\documentclass[a4paper,12pt]{report}
\usepackage{amsfonts}
\usepackage{amsmath}
\usepackage{graphics}
\usepackage{graphicx}

\usepackage{epstopdf}

\usepackage{atbegshi} 
\AtBeginDocument{\AtBeginShipoutNext{\AtBeginShipoutDiscard}} 

\author{\\Student: Holger Paul Keeler, BSc, BTech\\
\\Supervisor: Dr Scott McCue
\\
\\
\\
\\
\\
\\ Faculty of Science \\Griffith University\\
\\
\\
\\
\\
Submitted in partial fulfillment of the requirements of \\the
degree of Bachelor of Science with Honours\\
\vspace{1em}
\large{October 27, 2005}\\
\date{}
}
\title{\textbf{Free surface flow due to a submerged source}}

\newcommand{\cpvint}{\displaystyle\mbox{\boldmath$-$\kern-1.07em}\int}
\newcommand{\bm}[1]{\mbox{\boldmath $#1$}}
\newcommand{\sfrac}[2]{{\textstyle\frac{#1}{#2}}}

\begin{document}

\maketitle\newpage

\pagenumbering{roman}

\begin{LARGE}
\begin{center}
{\bf Abstract}
\end{center}
\end{LARGE}
In fluid mechanics a free surface problem entails a free moving
interface between one fluid (usually water) and another fluid
(usually air). Due to the unknown form of the interface and the
nonlinear boundary conditions applied there, free surface problems
are highly nonlinear and almost impossible to solve analytically.

In this project we consider the free surface flow due to a
submerged source in a channel of finite depth.  This problem has
been considered previously in the literature, with some
disagreement about whether or not a train of waves exist on the
free surface for Froude numbers less than unity.  The physical
assumptions behind the accepted model are clearly stated and
governing equations and boundary conditions derived. Complex
variable theory is then employed to obtain a singular nonlinear
integral equation, which describes the flow field completely.

The integral equation is evaluated numerically using a collocation
method, and implemented with mathematical software {\em Matlab}.
The numerical solutions suggest that indeed waves do exist on the
free surface, but are exponentially small in the limit that the
Froude number approaches zero.  An asymptotic expression is also
derived to solve the integral equation, valid for small Froude
numbers. This expansion is calculated to second order, which
improves on the leading order solution given previously in the
literature. Such a scheme is unable to predict waves on the free
surface, since they are exponentially small.  It is discussed how
exponential asymptotics could be used to derive a more accurate
analytic solution that describes waves on the free surface.

\newpage

\vspace*{30ex} This work has not previously been submitted for a
degree or diploma in any university. To the best of my knowledge
and belief, the dissertation contains no material previously
published or written by another person except where due reference
is made in the dissertation itself.

\begin{flushright}
\vspace{5ex}
---------------------------------------------
\\
Holger Paul Keeler \hspace{7ex}

\end{flushright}

\newpage

\tableofcontents

\listoffigures
\newpage

\pagenumbering{arabic}

\chapter{Introduction}
Historically, the behaviour of fluids, particularly water, has
been an extensively studied area in engineering, physics and
applied mathematics. For the simplest problems, much light has
been shed on this field over the years. However, when treating
more detailed and interesting cases the governing equations and
boundary conditions become increasingly complex and nonlinear.
Depending on their complexity and nature, fluid mechanic problems
can be divided into different classes and various analytical and
numerical solutions are developed accordingly. In this thesis we
investigate one interesting class of problems known as free
surface problems, which are highly nonlinear in nature, and must
be treated with a combination of analytical and numerical means.
In particular, we shall consider a steady state flow system due to
a source or sink submerged in a channel of fluid.

\section{Flow problems with free-surfaces}
The class of problems where the uppermost surface is the interface
between one fluid (usually water) and another fluid (usually air)
is known as free surface problems. This term stems from the fact
that the interface is \textit{free} to move with the evolution of
the system. The actual form of the free surface boundary is one of
the unknowns in these problems, hence determining the shape of the
free surface is part of the solution.

In general, free surfaces obey the Bernoulli principle, which
yields a nonlinear boundary condition along the free surface. This
nonlinear condition, coupled with the unknown form of the free
surface, rapidly reduces the tractability of free surface
problems, especially in terms of calculating analytical solutions.
In order to overcome this difficulty, numerous methods have been
employed to simplify free-surface problems. These methods have
often involved various approximations and numerical schemes.
Ideally, one wishes to employ a method that treats the free
surface problem analytically as far as possible, while only
resorting to numerical schemes in the final stages of the
solution, which naturally produces more accurate results.

One such method is to employ the feasible assumptions that the
flow is irrotational, and the fluid is incompressible and
inviscid, which is the approach used in this thesis. These
assumptions lead to the system obeying Laplace's equation
everywhere within the domain except in the neighbourhood of
singularities such as sources and sinks. This allows
two-dimensional problems to be treated analytically with complex
analysis and boundary integral methods, whilst numerical methods
are only employed to solve the resultant integral equations.

\section{Flow due to a submerged source or sink}\label{sect:flow}
One particular problem of interest is that of a steady flow system
due to a line source or sink in a channel of fluid of finite depth
and infinite length. Hocking \cite{H1995} discusses how these flow
models can be used to study the pumping and draining of various
bodies of water such as reservoirs, cooling ponds or solar ponds,
which usually do not consist of homogeneous fluid but rather a
number of stratified layers. The fluid stratification is a direct
result from the varying densities of the fluid, which arise from
temperature gradients and contaminant concentrations.

Particularly in reservoirs, a layer of fresh water usually overlays
a deeper layer, which may contain unwanted salts and pollutants.
As a result, it is often imperative that management carefully
controls the rate of water withdrawal in such a manner that the
neighbouring fluid is not drained \cite{HF1997}. Such controlled
procedures may be used, for example, to remove a lower layer of
polluted water in a reservoir, and leave the remaining
uncontaminated water. In addition, controlled withdrawal and
inflow of solar ponds may be used to manage the fluid
stratification, which stabilizes and optimizes the system
\cite{H1995}.

The solutions of such free surface problems also has a scientific
interest as many rudimentary questions in this area remain
unanswered. Finally, it should be noted that similar fluid flow
models can be used to describe other flow systems such as flow
under a boat bow \cite{T1991} or a sluice gate \cite{V1997}, and
accordingly, similar conformal mapping and numerical methods are
applicable for these related problems.  We discuss the
relationship between the problem of flow due to a source or sink
with the related problems further in
Section~\ref{sec:relatedfsproblems}.

We are interested in the steady state form of the free surface, with particular interest
in whether or not wave formation occurs on the free
surface when the system is under the effect of gravity. The flow
of such systems is characterized by the Froude number, $F$,
defined in general as
$$
F=\frac{m}{\sqrt{gL^3}},
$$
where $m$ is the flux in the far field due to the source, $g$ is
the acceleration due to gravity and $L$ is a typical length of the
system. The value for the typical length differs depending on
which Froude number is desired. The Froude number which uses the
average height of the fluid in the far field as the typical length
is known as the depth-based Froude number, $F_B$. One important
factor of this characteristic number is its ability to predict
whether or not wave-formation is possible for this type of
problem.

It can be shown (see Stoker~\cite{STOKER} or
Debnath~\cite{DEBNATH}, for example) for flow systems that waves
cannot occur on free surfaces when $F_B > 1$  while waves can
occur when $F_B < 1$. These types of flow systems are known as
\textit{supercritical} and \textit{subcritical}, respectively. It
should be noted that while this terminology originates from the
linear theory of water waves, their definitions have evolved to
described nonlinear wave phenomena.

The problem presented here is that of a subcritical system, which
has gained some interest in recent years. One early solution of
note was obtained by Hocking and Forbes \cite{HF1992}. This
solution laid down the foundation for the analysis to be performed
here by using a classic boundary integral method in a conformally
mapped plane to solve a system with low Froude numbers. This solution, however, failed to predict waves of
any size on the free surface.

During the same year, Mekias and Vanden-Broeck \cite{MV1991} used
a slightly different method, which involved reflecting the system
along the channel bottom before applying the boundary integral.
Their results disagreed with those of Hocking and Forbes
\cite{HF1992} by showing that a train of waves did actually occur
upstream on the free surface. The numerical results suggested that
the amplitude of the waves became exponentially small as $F_B$
approached zero.

Vanden-Broeck \cite{V1996} attempted to reconcile the conflicting
results by employing Hocking and Forbes' solution method to solve
the problem. The results agreed with that of Mekias and
Vanden-Broeck \cite{MV1991} by showing a train of waves exists on
the free surface. Additionally, it was shown that for lower Froude
numbers, the waves were so small that the free surface was
effectively flat, agreeing with the profiles of Hocking and Forbes
\cite{HF1992}.

Finally, Hocking and Forbes \cite{HF2000} employed another
boundary integral method similar to that of Hocking and Forbes
\cite{HF1992} but with surface tension taken under consideration.
While, surface tension shall be neglected here, the results also
showed that waves were possible on the free surface. This type of
solution is of interest as it offers a viable solution method in
solving the problem presented here.

The well-established methods \cite{Thomson} used here will follow
the lead of Hocking and Forbes solutions \cite{HF1992,HF2000} and
Vanden-Broeck \cite{V1996}. The solution methods
all involved conformal mapping and boundary integral methods in
order to produce a more tractable problem. The final nonlinear
equations obtained will need to be solved numerically with a
nonlinear minimization methods such as Newton-Raphson or
Levenberg-Marquardt scheme.

\section{Goals of this project}
This project is concerned with the free surface flow that arises
if a line source is placed beneath the surface of a fluid in a
channel of finite-depth.  This problem has been considered
previously in the literature by Hocking and Forbes \cite{HF1992},
Mekias and Vanden-Broeck \cite{MV1991}, Vanden-Broeck \cite{V1996}
and Hocking and Forbes \cite{HF2000}.  In particular, we wish to
achieve the following goals:
\begin{itemize}
\item We shall work through the formulation of the problem, giving a clear account of the
assumptions made, and their consequences.  This work shall be presented in Chapter~\ref{chap:mathform}.
\item The problem will be reformulated in Chapter~\ref{chap:bim} using a series of conformal maps following the
work of \cite{HF1992}, \cite{V1996} and \cite{HF2000}.  The entire nonlinear free boundary problem will be
reduced to solving a nonlinear singular integral equation.
\item A number of small but significant errors exist in the paper by Hocking and Forbes \cite{HF2000}.
One of our goals is to rectify these errors and clarify any misconceptions that may exist, and to
present a correct, readable and lucid account of the problem formulation and solution.
\item We shall solve the singular integral equation using a numerical scheme analogous to
\cite{HF1992}, \cite{V1996} and \cite{HF2000}.  This is already a
significant challenge given the nonlinear nature of the problem,
however much of the detail is left out of these papers, and as
such there are a number of issues that need clarifying and
amending. This part of the project will be detailed in
Chapter~\ref{chap:numscheme}.
\item Once the code is working the goal is to run the program a number of times using
various Froude numbers in order to vindicate the results of the Vanden-Broeck \cite{MV1991, V1996} by
showing the numerical existence of a wave-train upstream on the free surface.
Furthermore, we wish use the numerical scheme to derive a relationship between wave amplitude and the upstream Froude
number, and confirm that the amplitude of the waves is exponentially small in the limit
of small Froude number.  These results will be presented in Chapter~\ref{chap:results},
with the code given in Appendix~\ref{chap:code}.
\item It is possible to use a regular perturbation scheme to derive an asymptotic
solution in the limit that the Froude numbers approaches zero.  This approach was used by
Hocking and Forbes~\cite{HF1992}, who calculate the leading order term only.  One of the
goals of the project is to confirm the leading order solution of \cite{HF1992}, and to
extend this result by calculating the correction term.  We include this calculation in
Chapter~\ref{chap:asymptotic}.
\item The question of whether or not waves occur on the free surface is very important
for a number of reasons, as will discussed in Chapter~\ref{chap:conclude}.  A long term
goal for researchers in this area is to use asymptotic analysis to prove that waves always
exist on the free surface for the class of problems considered in this project, no matter
how small the Froude number is.  Such an analysis must be capable of capturing
exponentially small terms, unlike the perturbation scheme given by Hocking and
Forbes~\cite{HF1992} (and extended in Chapter~\ref{chap:asymptotic}).  It is likely that
these terms will be {\em beyond all orders}, and thus will require {\em exponential
asymptotics}, a highly sophisticated theory which has made enormous advances in recent
years (see Boyd~\cite{B1999}).  We shall discuss this possibility and outline the how
exponential asymptotics could be used in the present problem in
Chapters~\ref{chap:asymptotic} and \ref{chap:conclude}.
\item Finally, a goal is to put the problem of free surface flow due to a line source
into a greater context by relating it to other problems in hydrodynamics such as flow
past the bow or stern of a ship and flow past a sluice gate in Chapter
\ref{chap:conclude}.
\end{itemize}

\chapter{Mathematical Formulation}\label{chap:mathform}
In this chapter we shall outline our physical assumptions, which
have been made to simplify the problem of a flow system and to
present a more tractable mathematical model. The governing
equations and boundary conditions of the model shall be developed.
Finally, we shall nondimensionalize all variables, a process that
will reveal the number of independent parameters in the problem.

\section{Physical assumptions}
Our flow system consists of a layer of water with a layer of air
above it. It is assumed that the viscosity of the water can be
neglected thus the water is considered to be {\em inviscid}. This
is effectively equivalent to saying the viscous forces are much
smaller than the inertial forces, or that the {\em Reynolds
number} is large, which is certainly the case for withdrawal
problems. We assume that density of the water is constant and thus
the fluid is {\em incompressible}. Although in reality water is
compressible to a certain extent, its density variation on the
scale used in withdrawal problems is negligible. Finally, we
suppose that the flow system has started from rest at some initial
time, so that initially the vorticity of the fluid was zero. Since
the vorticity in an inviscid, incompressible fluid in constant
along a streamline, if it is initially zero then it is always so.
Hence the flow is considered to be {\em irrotational}. All of
these assumptions are standard, and are explained in further
detail in classic texts such as Batchelor~\cite{BATCHELOR} and
Milne-Thomson~\cite{Thomson}.

\section{Problem description}
The steady system consists of the irrotational flow of an
inviscid, incompressible fluid in a finite-depth channel, under
the effect of gravity, $g$. The symmetrical flow-domain is bounded
above by a free surface and below by a horizontal bottom. A line
source of strength $2m$ (that is $m$ is the flux in the far field)
is located beneath the surface. It is assumed that directly above
the submerged source on the free surface a stagnation point exists
where the fluid velocity is zero. This assumption is based on the
perfect symmetry of the system.

\begin{figure}[h]
\begin{center}
\includegraphics[height=6.5cm,width=10cm]{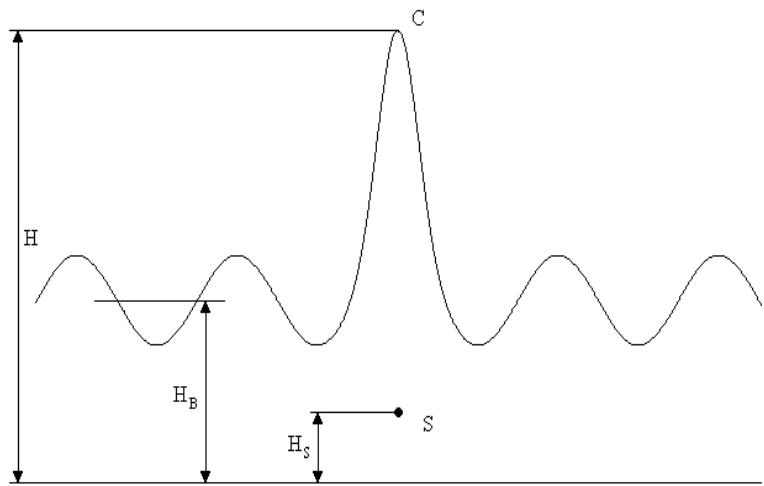}
\caption{2-D
profile of stagnation point with submerged source.\label{problem}}
\end{center}
\end{figure}

Due to a quadratic source term that arises later in the dynamic
boundary condition, the solution methods outlined here should
apply equally to problems involving sources or sinks. However,
there is an important exception. If the flow is due to a sink,
then it is not physically possible for a train of waves to exist
on the free surface in the far field. This intuitive requirement,
known as the {\em Sommerfield radiation condition}, can be derived
mathematically by showing that the group velocity for water waves
is always smaller than the phase velocity (see
Stoker~\cite{STOKER} or Debnath~\cite{DEBNATH}, say). Henceforth,
we shall refer to the singularity as a source while taking note
that the treatment applies similarly to sinks as well when the
Sommerfield radiation condition is satisfied.

In a similar manner to Vanden-Broeck \cite{V1996}, the physical
plane has its origin located at the stagnation point $C$, and is
located at a height of $H$ from the channel bottom. The average
height of the free surface upstream and the height of the source
are denoted by $H_B$ and $H_S$ respectively.

\section{Governing equations}
In nondimensional variables, we introduce the fluid velocity
vector, $\textbf{q}=u \textbf{i}+v\textbf{j}  $, which is related
to the vorticity vector, $\bm{\omega}$, by the curl expression
\begin{equation}\label{vort}
\nabla\times\textbf{q}=\bm{\omega}.
\end{equation}
The vorticity of the system is a measurement of the rate of
rotation of a fluid particle about the axes (Batchelor~\cite{BATCHELOR}). Since
we assume that the flow system is irrotational the vorticity
vanishes, which reduces equation (\ref{vort}) to
\begin{equation}\label{vortzero}
\nabla\times\textbf{q}=0.
\end{equation}
This result implies that the velocity vector can be written as
\begin{equation}\label{qphi}
\textbf{q}=\nabla\phi,
\end{equation}
where $\phi$ is known as the velocity potential of the system
\cite{BOAS}. From vector calculus, the velocity potential along
path $S$ is expressed as
\begin{equation}\label{phi}
\phi=\int_S\textbf{q}\cdot\textbf{dr}.
\end{equation}

We safely assume that the systems obeys the conservation of matter
which corresponds to a zero divergence
$$
\nabla\cdot\textbf{q}= 0,
$$
which together with equation (\ref{qphi}) leads to the linear
Laplace equation, namely
\begin{equation}\label{lap1}
\nabla\cdot(\nabla\phi)=\nabla^2\phi=0.
\end{equation}
It is well-established that working with the scalar quantities
such as $\phi$, instead of vector quantities, markedly simplifies
the solution method \cite{Thomson}. Hence, this approach has been
employed here.

\section{Kinematic conditions}
Naturally, the flow of fluid does not cross the free surface or
channel bottom. Since the fluid must flow along the boundaries,
the angle of the flow between the horizon, $\delta$, is simply the
angle of the fluid velocity $\textbf{q}$ along these streamlines.
This angle is obtained in a straightforward manner by using the
relation
$$\tan(\delta)=\frac{\bar{v}}{\bar{u}},$$
where $\bar{u}$ and $\bar{v}$ are the respective horizontal and
vertical components of the fluid velocity. Note that the bars
denote dimensional variables. The angle of the fluid velocity
along the streamlines is just a measurement of the gradient,
$$\tan(\delta)=\frac{d\bar{y}}{d\bar{x}}$$
of any boundary line. Additionally, dimensional Cartesian
variables $\bar{x}$ and $\bar{y}$ have been introduced and bars
have been employed to signify which variables have dimensions.
These relations lead to the following kinematic condition being
held along all boundaries,
\begin{equation}\label{kinematic}
\bar{v}=\bar{u}\frac{d\bar{y}}{d\bar{x}}.
\end{equation}

Along the channel bottom the vertical component of the fluid
velocity is zero, which yields the following kinematic condition

\begin{equation}\label{kinematic2}
\begin{array}{cc}
  \bar{v}=0, & \bar{y}=-H.
\end{array}
\end{equation}

\section{Dynamic conditions}
For a steady flow of an inviscid fluid under the effect of
gravity, the well-established Bernoulli equation \cite{BATCHELOR}
describes the system
\begin{equation}\label{dbernoulli1}
\frac{1}{2}\bar{q}^2+\frac{p}{\rho}+g\bar{y}=K,
\end{equation}
where $p$ and $\rho$ are the fluid pressure and density
respectively, and $K$ is some constant of the system. We have
assumed that the air pressure exerted upon the free surface of the water is
constant, which is not strictly true in reality but the minute
pressure variation on this scale can be neglected. The nonlinear
condition (\ref{dbernoulli1}) is held everywhere within the flow
domain. The fluid velocity at the stagnation point is naturally
equal to zero, thus if we apply the Bernoulli equation at the
stagnation point (that is, on the free surface where $\bar{x}=0,
\bar{y}=0$), the equation reduces to the following fluid-surface
condition
$$\frac{p_A}{\rho}=K,$$
where $p_A$ is the atmospheric pressure experienced anywhere on
the free surface. This condition states that the pressure along
the entire free surface is constant. By substituting this result
into the original equation (\ref{dbernoulli1}), the  Bernoulli
equation for anywhere along the free surface is expressed as
\begin{equation}\label{dbernoulli2}
\frac{1}{2}\bar{q}^2+g\bar{y}=0.
\end{equation}

\section{Nondimensionalizing}
For mathematical treatment, system variables need to be
nondimensionalized. Distances and velocities shall be
nondimensionalized with respect to $H$ and $m/H$, respectively.
The bars on all nondimensionalized variables shall be removed to
show that they are indeed dimensionless. In this fashion, the
Cartesian coordinates, the fluid velocity components and the fluid
velocity, respectively, become
$$x=\frac{\bar{x}}{H}, \quad y=\frac{\bar{y}}{H},\quad u=\frac{\bar{u}H}{m},
\quad v=\frac{\bar{v}H}{m},\quad q=\frac{\bar{q}H}{m}.$$ The
average upstream depth and the height of the source are denoted by
$\gamma$ and $\mu$ respectively. In nondimensional variables, the
flux in the far field and the height of the stagnation point are
both equal to unity.

The new dimensionless variables are substituted into equation
(\ref{dbernoulli2}) to yield
$$
\frac{1}{2}\left(\frac{qm}{H}\right)^2+gHy=0
$$
and the expression is divided through by $H$ and $g$ to obtain the
nondimensional Bernoulli equation
\begin{equation}\label{dbernoulli3}
\frac{1}{2}\left(\frac{m^2}{gH^3}\right)q^2+y=0,
\end{equation}
which is to be applied on the free surface.
\\
\begin{figure}[h]
\begin{center}
\includegraphics[height=6.5cm,width=10cm]{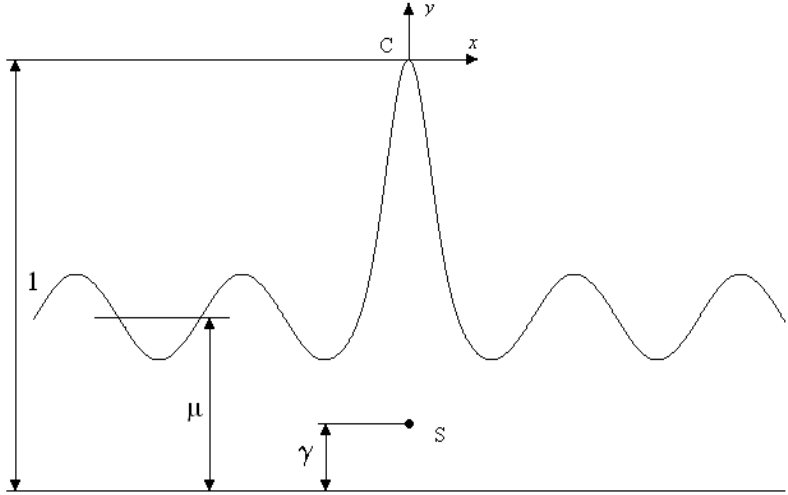}
\caption{Profile of the free surface problem with nondimensional
variables.\label{nonproblem}}
\end{center}
\end{figure}

\section{Froude numbers}
The depth-based Froude number was discussed earlier in regards to
its ability to predict the possibility of wave-formation. In its
complete form it is expressed as
\begin{equation}\label{fb}
F_B=\frac{m}{\sqrt{gH_B^3}}.
\end{equation}
The Froude number that uses the stagnation-point height as its
typical length is denoted by $F_{SP}$ and is similarly expressed
as
\begin{equation}\label{fsp}
F_{SP}=\frac{m}{\sqrt{gH^3}}.
\end{equation}
After some algebraic manipulation and the use of dimensionless
variables in equation (\ref{fb}), a relation between the two
Froude numbers is ascertained to be
\begin{equation}\label{fbb}
F_B=\frac{m}{\sqrt{gH_B^3}}=\frac{m}{\sqrt{gH}}\sqrt{\frac{H^3}{H_B^3}}=\frac{F_{SP}}{\mu^{3/2}}.
\end{equation}

Inspection of equation (\ref{dbernoulli3}) reveals that the
nondimensional term $\frac{m^2}{gH^3}$ is simply the square of
the stagnation-point Froude number, which reduces the Bernoulli
equation to
\begin{equation}\label{fspbernoulli}
\frac{1}{2}F_{SP}^2q^2+y=0,
\end{equation}
a condition satisfied along the free surface.

\section{Summary}
We have considered the problem of the flow due to a line source
with a stagnation point on the free surface.  The problem has been
nondimensionalized so that the stagnation point is located at the
origin, and the source is at the point $(x,y)=(0,-1+\gamma)$. The
flow is bounded below by a horizonal bottom at $y=-1$, and far
downstream the average level of the fluid is located at
$y=-1+\mu$.

The system can be effectively described by the nondimensional
quantity $\phi$, which is a velocity potential.  From our
assumptions that the fluid is incompressible and inviscid and that
the flow is irrotational, we established that $\phi$ satisfies
Laplace's equation
\begin{equation}
\nabla^2\phi=\frac{\partial^2\phi}{\partial x^2}+\frac{\partial^2\phi}{\partial y^2}=0,
\label{laplace}
\end{equation}
throughout the flow domain, except at the source, where it behaves
like
\begin{equation}
\phi\sim \frac{1}{2\pi}\log[x^2+(y+1-\gamma)^2]\quad\mbox{as}\quad
(x,y)\rightarrow(0,-1+\gamma). \label{source}
\end{equation}
On the horizonal bottom we have the boundary condition
\begin{equation}
\frac{\partial \phi}{\partial y}=0, \quad y=-1,
\label{bottom}
\end{equation}
while on the free surface we have the two boundary conditions
\begin{equation}
\frac{\partial \phi}{\partial y}=\frac{\partial \phi}{\partial x}\frac{dy}{dx},
\end{equation}
\begin{equation}
\frac{1}{2}F_{SP}^2\left|\nabla\phi\right|^2+y=0,
\label{dynamic}
\end{equation}
where the parameter $F_{SP}$ is the Froude number.  Note that the problem is
nonlinear because of the unknown free surface as well as the
nonlinear nature of the last boundary condition, Bernoulli's
equation.  There is, not surprisingly, no analytical solution to
this problem; however we can solve it using a combination of analytic and
numerical techniques, and these are described in the following
chapters.

\chapter{Boundary Integral Method}\label{chap:bim}
In the previous chapter we derived the governing equations for our
problem, including Laplace's equation and appropriate boundary
conditions pertaining to our physical assumptions of the flow
system. In the present chapter we shall formulate the problem with
the use of complex variable theory to describe our model, and
simplify the geometry with the use of conformal mapping. A
boundary integral method is then used to derive an integral
equation, which describes our problem completely.

\section{Complex variables}
We have established that the flow of the system presented here
obeys the equation $\textbf{q}=\nabla\phi$, within the flow domain
except at the source, which implies
\begin{equation}\label{cr1}
u=\frac{\partial\phi}{\partial x},\quad
v=\frac{\partial\phi}{\partial y}. \quad
\end{equation}
We introduce the quantity $\psi$, known as the stream function, defined by
\begin{equation}\label{cr2}
u=\frac{\partial\psi}{\partial y}, \quad
v=-\frac{\partial\psi}{\partial x}.
\end{equation}
We combine equations (\ref{cr1}) and (\ref{cr2}) to yield the
well-renowned Cauchy-Riemann equations \cite{RAHMAN}, namely
\begin{equation}\label{cr}
\frac{\partial\phi}{\partial x}=\frac{\partial\psi}{\partial y},
\quad \frac{\partial\phi}{\partial
y}=-\frac{\partial\psi}{\partial x}.
\end{equation}

We can introduce complex analysis and use a complex potential
function to describe the system \cite{Thomson}. We let the complex
variable $z=x+iy$ represent the physical plane and introduce the
complex potential
\begin{equation}\label{f}
f(z)=\phi(x,y)+i\psi(x,y),
\end{equation}
which, since $\phi$ and $\psi$ obey the Cauchy-Riemann equations,
is an analytic function of $z=x+iy$ everywhere within the flow
domain except at the source.

The line source is a mathematical singularity \cite{RAHMAN}, and
as a result, the complex potential $f(z)$ is not analytic there. In
the neighbourhood of the source the potential is given by
$$
f(z)\sim\frac{1}{\pi}\ln (z-z_0) \quad \textrm{as} \quad
z\rightarrow z_0,
$$
where $z_0$ marks the position of the source on the complex plane.
Using the nondimensional variables, the above expression becomes
\begin{equation}\label{fs}
f(z)\sim\frac{1}{\pi}\ln (z+1-\gamma) \quad \textrm{as} \quad
z\rightarrow \gamma-1.
\end{equation}

We introduce the complex conjugate velocity, $w$ expressed as
\begin{equation}\label{w}
f'(z)=w(z)=u-iv,
\end{equation}
which also is an analytic function.  The complex velocity is
related to the fluid velocity simply by
\begin{equation}\label{qw}
w=qe^{-i\delta}.
\end{equation}
where $\delta$ is the aforementioned angle of the velocity vector
in relation to the horizontal axis and $q=\sqrt{u^2+v^2}$ is the
magnitude of the velocity.

\section{Symmetry of the system}
It is been assumed that the problem presented here is perfectly
symmetric about the $y$-axis. This symmetry allows the problem to
be simplified by only treating one side of the system. The
right-hand side half is arbitrarily chosen for analysis, which
corresponds to the source strength becoming $m$.

The reflection process here is physically akin to inserting a
vertical wall of infinitesimal width directly in the centre of the
system thus creating two mirrored systems. Hence, we shall now
assume that a solid boundary exists to the left side. Naturally,
when solutions are obtained for the right-hand side, it is a
straightforward process of reflecting the results in order to gain
a physical description of the whole system.
\begin{figure}[h]
\begin{center}
\includegraphics[height=6.5cm,width=7cm]{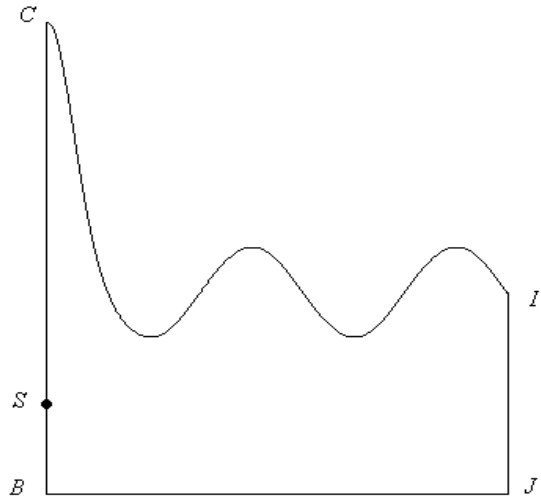}
\caption{Right-hand side of problem with a solid boundary located
to the left. \label{halfproblem}}
\end{center}
\end{figure}

\section{Potential function}\label{sec:potfun}
All known values of the potential function for the system need to
be plotted onto the $f$-plane. While maintaining the system's
generality, we let $\psi = 0$ on the free surface and $\phi = 0$
at the stagnation point, $C$.

With the $f$ values set for the stagnation point, we now
investigate the potential values along the channel bottom, in
particular directly beneath the source at point $B$. By
integrating equation (\ref{cr2}) up to the $x$-axis, we obtain a
general expression for the stream function
$$
\psi=-\int^0_{y}udy' + k(x) + c,
$$
where $y'$ is a dummy variable, $c$ and $k(x)$ is a constant and
function of $x$ due to integration. Since the stream function is
constant anywhere along the channel bottom, regardless of the $x$
value, the function $k(x)$ disappears. Intuitively, since the
stream function was set to zero at the origin, the integration
constant must be zero, which reduces the equation to
$$
\psi=-\int^0_{y}udy'.
$$
Upon inspection, we observe that the integral is simply the
negative of the flux across the defined region. From the
conservation of mass, we know that the total flux must be equal to
the dimensionless source strength, which is equal to one. Thus,
$\psi = -1$ anywhere along the channel bottom.

At point $B$, the value of $\phi$ is unknown and is denoted as
$\phi_B$, and is arbitrarily placed on the left-hand side of the $f$-plane.

\begin{figure}[h]
\begin{center}
\includegraphics[height=5.0cm,width=8cm]{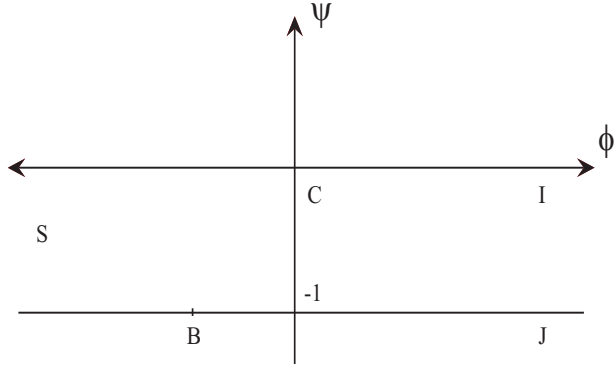}
\caption{The problem on the $f$-plane.\label{fplane}}
\end{center}
\end{figure}

The location of the source at point $C$ on the $f$-plane is
obtained with a physical argument. It is known that the source is
a singularity, and thus, the the velocity of fluid leaving the
source has effectively an infinite speed, which corresponds to
point $S$ being located at negative infinity on the $\phi$-axis.
Finally, all fluid in the system originates from the source, hence
all streamlines do as well. All streamlines spread out between the
free surface and the channel bottom. Since the stream function
values are known for the boundaries, intuitively the values for
$\psi$ at the source must range from $0$ to $-1$.

\section{Mapping the potential}
All the points of the system are now located on the infinite strip
between $\psi=0$ and $\psi=-1$ on the $f$-plane. In order to
reformulate this problem using a boundary integral method we
introduce the complex variable $\zeta=\xi+i\eta$, allowing the
infinite strip to be mapped onto the lower-half of the
$\zeta$-plane with the conformal transformation
\begin{equation}\label{map}
\zeta=\xi+i\eta=e^{\pi f}.
\end{equation}
Point C is located at $\phi=0$ and $\psi=0$ on the $f$-plane,
which corresponds to
$$\zeta=\xi_C=1.$$
Point B is located at $\phi=\phi_B$ and $\psi=-1$ on the
$f$-plane, which corresponds to
$$\zeta=\xi_B=-e^{\pi\phi_B}.$$
Point I is located at $\phi=\infty$ and $\psi=0$ on the $f$-plane,
which corresponds to
$$\zeta=\xi_I=\infty +0i.$$
Point J is located at $\phi=\infty$ and $\psi=-1$ on the
$f$-plane, which corresponds to
$$\zeta=\xi_J=-\infty +0i.$$
Point S is located at $\phi=-\infty$ and between $\psi=0$ and
$\psi=-1$ on the $f$-plane, which corresponds to
$$\zeta=\xi_S=0.$$
All points of importance are located on the real axis on the
$\zeta$-plane. The free surface corresponds to $1<\xi<\infty$
while the channel bottom corresponds to $-\infty<\xi<\xi_B$. The
wall is divided into two sections: beneath the source corresponds
to $\xi_B<\xi< 0$ while above the source corresponds to $0<\xi<1.$
\begin{figure}[h]
\begin{center}
\includegraphics[height=5.0cm,width=8cm]{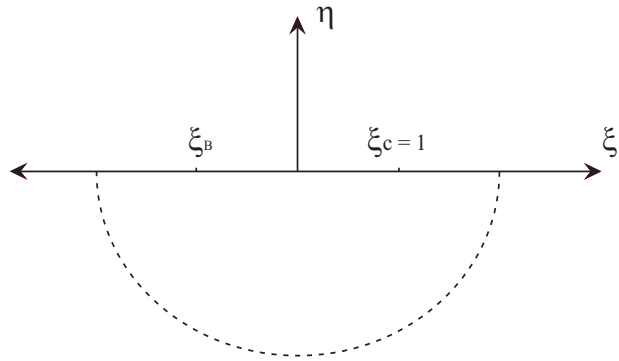}
\caption{The
problem mapped to the lower half of the $\zeta$-plane.
\label{zetaplane}}
\end{center}
\end{figure}

\section{Logarithmic hodograph}
The governing equations need to be reformulated for the
$\zeta$-plane with the geometry of the system embedded. We
introduce a new analytic function, known as the logarithmic
hodograph \cite{T1991}, defined as
\begin{equation}\label{hodo0}
\Omega(\zeta) = \delta(\zeta) + i\tau(\zeta),
\end{equation}
which is related to the complex velocity (\ref{w}) by
\begin{equation}\label{hodo1}
f'(z(\zeta))=u-iv=\frac{1}{\mu}e^{-i\Omega(\zeta)}
\end{equation}
where $\mu$ is the nondimensional average depth of the fluid
upstream. In order to gain some physical interpretation of the
function, we recast it in the following form
\begin{equation}\label{hodo2}
w=u-iv=\frac{1}{\mu}e^{\tau(\zeta)}e^{-i\delta(\zeta)}.
\end{equation}
By comparing result (\ref{hodo2}) to equation (\ref{qw}), we
observe that $q=\frac{1}{\mu}e^{\tau(\zeta)}$ and
$\delta=\delta(\zeta)$. Thus, $\frac{1}{\mu}e^{\tau(\zeta)}$ gives
the magnitude of the fluid velocity and $\delta(\zeta)$ gives the
angle between the streamlines and the $x$-axis.

\section{Boundary values}
As previously discussed, the kinematic conditions ensure that the
fluid flow follows the boundaries of the system. Hence, the
physical interpretation of $\delta(\zeta)$ along the boundaries is the actual angle of the
the solid boundaries to the horizonal.  Further, along the boundaries the complex
variable $\zeta$ takes real values $\xi$.  Thus, by inspecting the orientation of
the different boundaries on the physical plane, the values of
$\delta (\xi)$ are obtained for all the $\xi$ values that
correspond to solid boundaries, as follows:
\begin{equation}\label{delta}
\delta (\xi) = \left\{ \begin{array}{ccc}
  0 & \textmd{if} & -\infty < \xi < \xi_B, \\
  -\pi/2 & \textmd{if} & \xi_B< \xi < 0, \\
  \pi/2 & \textmd{if} & 0 < \xi < 1, \\
  \textmd{unknown} & \textmd{if} & 1 < \xi < \infty.
\end{array}\right.
\end{equation}
\begin{figure}[h]
\begin{center}
\includegraphics[height=5.0cm,width=8cm]{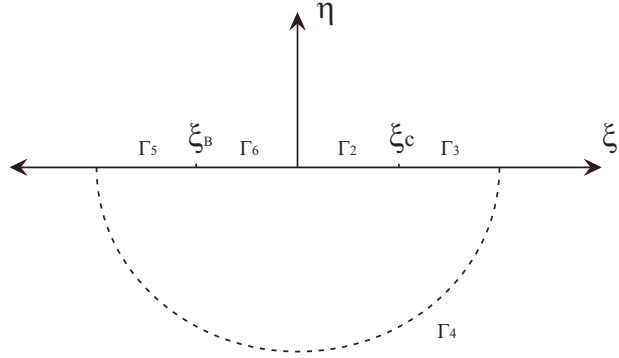}
 \caption{The semi-circle $\Gamma$ separated into different regions on the
$\zeta$-plane. \label{gamma}}
\end{center}
\end{figure}

\section{Cauchy's Theorem}
An equation that relates $\tau(\xi)$ and $\delta(\xi)$ on the real axis can be
ascertained by applying a circular integral of the function
$\Omega(\zeta)$ around the curve denoted by $\Gamma$ in an
anti-clockwise direction. With the singularity on the exterior of
the integral region, we invoke the famous Cauchy integral theorem
\cite{Arfken}, namely
\begin{equation}\label{cauchy}
\oint_{\Gamma}{\frac{\Omega(\zeta)}{\zeta-\zeta_0}} d\zeta=0,
\end{equation}
where $\zeta_0$ is a point outside $\Gamma$.  In fact we let $\zeta_0$ lie on the real
axis.  The singular point $\zeta_0=\xi_0$ is avoided by forming a detour in the shape of a
semi-circle of vanishing radius around it denoted by $\Gamma_1$.
We use the residue theorem \cite{BOAS} to obtain the integral
contribution of the $\Gamma_1$, namely
$$
\int_{\Gamma_1}{\frac{\Omega(\zeta)}{\zeta-\xi_0}}
d\zeta=i\pi\Omega(\xi_0).
$$
Also, the integral contribution of the semi-circle of
infinite radius, denoted by $\Gamma_4$, yields
$$
\int_{\Gamma_4}{\frac{\Omega(\zeta)}{\zeta-\xi_0}} d\zeta=0.
$$
These results are substituted into equation (\ref{cauchy}) to give
\begin{equation}\label{cauchy2}
\Omega(\xi_0)=\frac{i}{\pi}\cpvint_{-\infty}^{\infty}{\frac{\Omega(\xi)}{\xi-\xi_0}}d
\xi ,
\end{equation}
which by taking the real and imaginary parts give
\begin{equation}\label{taubar}
\tau(\xi_0)=\frac{1}{\pi}\cpvint_{-\infty}^{\infty}{\frac{\delta(\zeta)}{\xi-\xi_0}}d
\xi,
\end{equation}
\begin{equation}\label{deltabar}
\delta(\xi_0)=-\frac{1}{\pi}\cpvint_{-\infty}^{\infty}{\frac{\tau(\xi)}{\xi-\xi_0}}d
\xi,
\end{equation}
where the bars denote Cauchy Principal Values. The integral region
in equation (\ref{taubar}) is broken into separate integral
regions marked as follows:
$$
\begin{array}{cccc}
  \textrm{Region}& & \textrm{$\xi$ Lower Limit}& \textrm{ $\xi$ Upper Limit} \\
  \textrm{II} & &   0 &1\\
  \textrm{III} & &   1 &\infty \\
  \textrm{V} & &-\infty &\xi_B \\
  \textrm{VI} & &\xi_B&0
\end{array}
$$
Thus, the integral in equation (\ref{taubar}) is broken into four
integrals
\begin{eqnarray*}
\tau(\xi_0) & = &
\int_{\Gamma_2}{\frac{\delta(\xi)}{\xi-\xi_0}} d\xi
+\int_{\Gamma_3}{\frac{\delta(\xi)}{\xi-\xi_0}} d\xi
+\int_{\Gamma_5}{\frac{\delta(\xi)}{\xi-\xi_0}} d\xi
+\int_{\Gamma_6}{\frac{\delta(\xi)}{\xi-\xi_0}} d\xi \\
& = & \frac{1}{\pi}\int^{1}_{0}{\frac{\delta(\xi)}{\xi-\xi_0}} d\xi
+\frac{1}{\pi}\cpvint^{\infty}_{1}{\frac{\delta(\xi)}{\xi-\xi_0}}
d\xi
+\frac{1}{\pi}\int^{\xi_B}_{-\infty}{\frac{\delta(\xi)}{\xi-\xi_0}}
d\xi+\frac{1}{\pi}\int^{0}_{\xi_B}{\frac{\delta(\xi)}{\xi-\xi_0}}
d\xi.
\end{eqnarray*}
The $\delta(\xi)$ values  are known for regions II, V, and VI from
the flow conditions (\ref{delta}).  Substitution of these values gives
\begin{equation}\label{intc}
\tau(\xi_0)=
\frac{1}{\pi}\int^{1}_{0}{\frac{\pi/2}{\xi-\xi_0}} d\xi
+\frac{1}{\pi}\cpvint^{\infty}_{1}{\frac{\delta(\xi)}{\xi-\xi_0}}
d\xi +\frac{1}{\pi}\int^{0}_{\xi_B}{\frac{-\pi/2}{\xi-\xi_0}}
d\xi,
\end{equation}
leaving the remaining two integrals to be evaluated. The integral
for region II yields
\begin{equation}\label{intIIa}
\frac{1}{\pi}\int^{1}_{0}\frac{\pi}{2}{\frac{1}{\xi-\xi_0}}
d\xi =\frac{1}{2}\ln\left(\frac{1-\xi_0}{-\xi_0}\right),
\end{equation}
while the integral for region VI yields
\begin{equation}\label{intVIa}
\frac{1}{\pi}\int^{0}_{\xi_B}\frac{-\pi}{2}{\frac{1}{\xi-\xi_0}}
d\xi =\frac{1}{2}\ln\left(\frac{\xi_B-\xi_0}{-\xi_0}\right).
\end{equation}
Addition of results (\ref{intVIa}) and (\ref{intIIa}) gives
$$
\frac{1}{\pi}\int^{1}_{0}{\frac{\delta(\xi)}{\xi-\xi_0}} d\xi
+\frac{1}{\pi}\int^{0}_{\xi_B}{\frac{\delta(\xi)}{\xi-\xi_0}} d\xi
=\frac{1}{2}\ln\left[\frac{(1-\xi_0)(\xi_B-\xi_0)}{\xi_0^2}\right]
$$
This result is substituted into the original integral (\ref{intc})
to obtain the expression
$$
\tau(\xi_0)=\frac{1}{2}\ln\left[\frac{(\xi_0-1)(\xi_0-\xi_B)}{\xi_0^2}\right]
+\frac{1}{\pi}\cpvint^{\infty}_{1}{\frac{\delta(\xi)}{\xi-\xi_0}}
d\xi.
$$
The variables are aptly changed to obtain the equation
\begin{equation}\label{tau1}
\tau(\xi)=\frac{1}{2}\ln\left[\frac{(\xi-1)(\xi-\xi_B)}{\xi^2}\right]
+\frac{1}{\pi}\cpvint^{\infty}_{1}{\frac{\delta(\xi')}{\xi'-\xi}}
d\xi', \quad \quad \xi>1.
\end{equation}
This equation effectively describes the magnitude of the fluid
velocity for a point located anywhere along the free surface
corresponding to $1<\xi<\infty$ on the $\xi$-axis.

Equation (\ref{tau1}) is valid for $\xi>1$ and as such gives real values
in this $\xi$ range. To calculate $\tau(\xi)$ values for $\xi_B<\xi<1$
we need to modify equation (\ref{tau1}) slightly.  When $\xi_B<\xi<1$
equation (\ref{tau1}) gives complex values due to its log term, or
more specifically the square root in the log term. To rectify this
we factor out and completely remove a negative one from the the
log term
$$
\frac{1}{2}\ln\left[\frac{(\xi-1)(\xi-\xi_B)}{\xi^2}\right]=
i\pi+\frac{1}{2}\ln\left[\frac{(1-\xi)(\xi-\xi_B)}{\xi^2}\right]
$$
Recall that one of the steps in deriving equation (\ref{taubar})
involved taking the imaginary part of the equation. Inspection of
the equation (\ref{tau1}) derivation reveals that imaginary the
term $i\pi$ in the above expression becomes real during the
derivation. This implies that when the imaginary part is taken to
yield an equation (\ref{tau1}), the $\pi$ term is not in equation
(\ref{tau1}). Thus, we can write equation (\ref{tau1}) as
\begin{equation}\label{tauz}
\tau(\xi)=\frac{1}{2}\ln\left[\frac{(1-\xi)(\xi-\xi_B)}{\xi^2}\right]
+\frac{1}{\pi}\int^{\infty}_{1}{\frac{\delta(\xi')}{\xi'-\xi}}
d\xi', \quad \quad \xi_B<\xi<1,
\end{equation}which is valid since it produces real values. Finally, the
singularity of the integral term has been removed due to the
change of $\xi$ value range, hence the Cauchy principal integral
becomes a regular integral.

\section{Mapping back to the physical plane}
Expressions for obtaining the $z$-plane points that lie on the
free surface are needed. In order to obtain these expressions, we
ascertain an expression for $z$ by rearranging equation
(\ref{hodo1}) to give
\begin{equation}\label{zvalue}
dz=\mu e^{i\Omega(\zeta)}df.
\end{equation}
By differentiating and rearranging the transformation (\ref{map})
we obtain the differential equation
$$
\frac{d\zeta}{df}=\pi e^{\pi f}=\pi\zeta,
$$
and hence equation (\ref{zvalue}) becomes
$$
dz=\mu e^{i\Omega(\zeta)}df=
dz=\frac{\mu}{\pi}\frac{e^{i\Omega(\zeta)}}{\zeta} d \zeta.
$$
On the boundaries in the physical plane, $\zeta$ is real so the
equation reduces to
$$
dz=\frac{\mu}{\pi}\frac{e^{i\Omega(\xi)}}{\xi} d \xi.
$$
This equation is integrated to yield an expression for $z$, namely
\begin{equation}\label{zvalue1}
z(\xi)=\frac{\mu}{\pi}\int_1^{\xi}\frac{e^{i\Omega(\xi')}}{\xi'} d
\xi'=\frac{\mu}{\pi}\int_1^{\xi}\frac{e^{-\tau(\xi')+i\delta(\xi')}}{\xi'}
d \xi'.
\end{equation}
The identity $e^{i \delta} = \cos \delta + i \sin \delta$ and
equation (\ref{hodo0}) are used to obtain respective expressions
for $x$ and $y$. By taking the real component of equation
(\ref{zvalue1}) we obtain the $x$ expression
\begin{equation}\label{xvalue}
x(\xi)=\frac{\mu}{\pi}\int_1^{\xi}\frac{e^{-\tau(\xi')}\cos
\delta(\xi')}{\xi'} d \xi'.
\end{equation}
In a similar fashion an expression for $y$ is obtained by taking
the imaginary component of equation (\ref{zvalue1}) to obtain
\begin{equation}\label{yvalue}
y(\xi)=\frac{\mu}{\pi}\int_1^{\xi}\frac{e^{-\tau(\xi')}\sin
\delta(\xi')}{\xi'} d \xi'.
\end{equation}

\section{Bernoulli's Equation on the $\zeta$-plane}
Inspection of the dimensionless Bernoulli equation
(\ref{fspbernoulli}) reveals that we need an expression for the
quantity $q^2$. From equation (\ref{hodo2}), we know that
magnitude of the complex velocity is equal to the magnitude of the
fluid velocity, namely
$$
q=\frac{1}{\mu}e^{\tau(\xi)},
$$
which is simply squared to yield the expression
\begin{equation}\label{qsquare}
q^2=\frac{e^{2\tau(\xi)}}{\mu^2}.
\end{equation}
Equations (\ref{qsquare}) and (\ref{yvalue}) are substituted into
equation (\ref{fspbernoulli}) to yield the Bernoulli equation for
the $\zeta$-plane, namely
\begin{equation}\label{zetabernoulli}
\frac{1}{2}F_{SP}^2 \frac{e^{2\tau(\xi)}}{\mu^2}
+\frac{\mu}{\pi}\int_1^{\xi}\frac{e^{-\tau(\xi')}\sin
\delta(\xi')}{\xi'} d \xi'=0,
\end{equation}
which applies for $\xi>1$, corresponding to the free surface.

\section{Nonlinear integral equation}
A second equation for $\tau(\xi)$, in addition to (\ref{tau1}), is
derived by differentiating equation (\ref{zetabernoulli}) with
respect to $\xi$ and rearranging the results to give
$$
 \frac{F_{SP}^2 }{\mu^2}\frac{d\tau(\xi)}{d\xi}e^{2\tau(\xi)}
+\frac{\mu}{\pi}\frac{e^{-\tau(\xi)}\sin \delta(\xi)}{\xi}=0,
$$
which simplifies to
$$
\frac{d\tau(\xi)}{d\xi}e^{3\tau(\xi)} +\frac{\mu^3}{\pi
F_{SP}^2}\frac{\sin \delta(\xi)}{\xi}=0.
$$
Some algebraic rearrangement and suitable integration is applied
to yield
$$
\int_1^{\xi}e^{3\tau(\xi')}d\tau(\xi') =-\frac{\mu^3}{\pi
F_{SP}^2}\int_1^{\xi}\frac{\sin \delta(\xi')}{\xi'}d\xi'.
$$
But the left-hand side of this equation is simply $\sfrac{1}{3}e^{3\tau(\xi)}$,
given that the fluid velocity approaches zero near the stagnation point,
implying that $e^{3\tau(\xi)}\rightarrow 0$ as $\xi\rightarrow 1$.  Thus
$$
e^{3\tau(\xi)}=-\frac{3\mu^3}{\pi
F_{SP}^2}\int_1^{\xi}\frac{\sin \delta(\xi')}{\xi'}d\xi',
$$
or
\begin{equation}\label{tau2}
\tau(\xi)=\frac{1}{3}\ln\left[-\frac{3\mu^3}{\pi
F_{SP}^2}\int_1^{\xi}\frac{\sin \delta(\xi')}{\xi'}d\xi'\right].
\end{equation}
Equations (\ref{tau1}) and (\ref{tau2}) are equated to yield a
singular nonlinear integral equation
\begin{equation}\label{tauf}
\frac{1}{2}\ln\left[\frac{(\xi-1)(\xi-\xi_B)}{\xi^2}\right]
+\frac{1}{\pi}\cpvint^{\infty}_{1}{\frac{\delta(\xi')}{\xi'-\xi}}
d\xi'=\frac{1}{3}\ln\left[-\frac{3\mu^3}{\pi
F_{SP}^2}\int_1^{\xi}\frac{\sin \delta(\xi')}{\xi'}d\xi'\right],
\end{equation}
which is valid for $\xi>1$, that is, anywhere along the free
surface.

The integral equation (\ref{tauf}) is a type of Nekrasov equation,
as discussed in Tuck~\cite{T1991} and Wehausen and
Laitone~\cite{WL}.  A solution to (\ref{tauf}) for $\delta(\xi)$
will automatically satisfy Laplace's equation (\ref{laplace}),
since Cauchy's theorem (\ref{cauchy}) works only if $f=\phi+i\psi$
is analytic.  The solution will satisfy the singular behaviour
(\ref{source}) near the source, as this has been taken care via
the conformal mapping.  The bottom condition (\ref{bottom}) and
the symmetry of the system is satisfied from substituting the
correct value for $\delta$ for $\xi<1$, as done in equations
(\ref{intIIa}) and (\ref{intVIa}).  The kinematic condition
(\ref{kinematic}) is identically satisfied by forcing $\psi=0$ on
the free surface, as discussed in Section~\ref{sec:potfun}.
Finally, Bernoulli's equation (\ref{dynamic}) is satisfied by any
solution to (\ref{tauf}), as described earlier in this section.

\chapter{Numerical Scheme}\label{chap:numscheme}
In the previous chapter we derived a singular nonlinear integral
equation which governs our problem of flow due to a submerged
source. After deriving two additional equations, we shall develop
a suitable method to represent and solve these equations for
$\delta(\xi)$ and $\mu$, which are then used to determine
$\tau(\xi)$.  We shall discuss how the singular nature of the
integral equations is handled and the briefly outline the details
of the numerical scheme. Once the values of $\delta(\xi)$, $\mu$
and $\tau(\xi)$ have been determined, it is a relatively
straightforward procedure of calculating the free surface profile
with equations (\ref{xvalue}) - (\ref{yvalue}).

\section{System equations}
In order to solve the unknown $\delta(\xi)$, $\tau(\xi)$  and
$\mu$ values with a suitable numerical scheme, the $\xi$-axis is
divided appropriately to yield $N$ mesh points. Since we want
these mesh points equally spaced in the physical plane for sufficiently large $x$, we use
evenly spaced mesh points in the potential function, $\phi$.
Hence, we introduce the change of variables
\begin{equation}\label{xiphi}
\xi=e^{\pi \phi}.
\end{equation}
The change of variables implies we can rewrite equations
(\ref{tau1}) and (\ref{tau2}) respectively as
\begin{equation}\label{tau1phi}
\tau(\phi)=\frac{1}{2}\ln\left[\frac{(e^{\pi \phi}-1)(e^{\pi
\phi}-e^{\pi \phi_B})}{e^{2\pi \phi}}\right]
+\cpvint^{\infty}_{0}{\frac{\delta(\phi')}{e^{\pi\phi'}-e^{\pi\phi}}}
e^{\pi\phi'}d\phi',
\end{equation}
\begin{equation}\label{tau2phi}
\tau(\phi)=\frac{1}{3}\ln\left[-\frac{3\mu^3}{
F_{SP}^2}\int_{0}^{\phi}\sin(\delta(\phi'))d\phi'\right]=0,
\end{equation}
which naturally transforms equation (\ref{tauf}) into
\begin{equation}\label{taufphi}
\frac{1}{2}\ln\left[\frac{(e^{\pi \phi}-1)(e^{\pi \phi}-e^{\pi
\phi_B})}{e^{2\pi \phi}}\right]
+\cpvint^{\infty}_{0}{\frac{\delta(\phi')e^{\pi\phi}}{e^{\pi\phi'}-e^{\pi\phi'}}}
d\phi'-\frac{1}{3}\ln\left[-\frac{3\mu^3}{
F_{SP}^2}\int_{0}^{\phi}\sin(\delta(\phi'))d\phi'\right]=0.
\end{equation}

We introduce the evenly spaced mesh points in $\phi$
\begin{equation}\label{phiI}
\phi_I=(I-1)\Delta\phi, \quad I=1,...,N,
\end{equation}
where $\Delta\phi$ is a constant spacing of $\phi$ and $N$ is the
number of points respectively. The corresponding unknowns are
written as
\begin{equation}\label{deltaI}
\delta_I=\delta(\phi_I), \quad I=1,...,N,
\end{equation}
\begin{equation}\label{tauI}
\tau_I=\tau(\phi_I), \quad I=1,...,N.
\end{equation}
It is assumed that the free surface is horizontally flat at the
stagnation point, hence $\delta_1=0$.  The height of the source above the bottom at point $B$ is an
input, and is governed by the value of $\xi_B$ (note that $\xi_B=0$ corresponds to a
source on the channel bottom).  Thus, there are $N$
unknowns, which are the $\mu$ and $(N-1)$ values of $\delta_I$, to
solve.

The integral equation (\ref{taufphi}) is applied at all points
between the second and the penultimate point inclusively along the
$\xi$-axis. The integral is evaluated with a simple trapezoidal
method, which yields $N-2$ nonlinear equations.

A further equation is obtained by slightly modifying equation
(\ref{yvalue}), substituting the nondimensional stagnation point
height and integrating from $\xi=\xi_B$ to $\xi=\xi_C=1$, which
corresponds to the bottom and top of the wall in the physical
plane, in order to produce the equation
\begin{equation}\label{y1}
1+\frac{\mu}{\pi}\int^1_{\xi_B}\frac{e^{-\tau(\xi')}\sin
\delta(\xi')}{\xi'} d \xi'=0,
\end{equation}
where the known flow conditions along the wall give $\delta(\xi)$
values for this $\xi$ range. Care must taken due to the singular
nature of the integrals and this is discussed in the next section.

The final equation is obtained by using a standard polynomial
extrapolation method~\cite{NUMR} to relate the last three delta
values, namely
\begin{equation}\label{extrap}
\delta_N=3\delta_{N-1}-3\delta_{N-2}+\delta_{N-3}.
\end{equation}

Finally, the $\xi_B$ value is a system input and has a value of
zero when the source is located on the channel floor. We now have
a closed system with $N$ equations and $N$ unknowns to solve.

\section{Singularities}
Our derived system of mostly integral equations has a number of
singularities that must be treated with care when applying the
trapezoidal scheme. It will be shown, by the use of appropriate
change of variables, that even though some integrands contain
singularities, the integrals are in fact finite in value.

Firstly we must deal the the Cauchy Principle Value integral in
(\ref{taufphi}). Following Hocking and
Vanden-Broeck~\cite{HV1997}, we do this by adding and subtracting
the singularity as
$$
\cpvint^{\infty}_{0}{\frac{\delta(\phi')e^{\pi \phi'}}{e^{\pi
\phi'}-e^{\pi \phi}}}
d\phi'=\int^{\infty}_{0}{\frac{\delta(\phi')-\delta(\phi)}{e^{\pi
\phi'}-e^{\pi \phi}}}e^{\pi \phi'}
d\phi'+\int^{\infty}_{0}{\frac{\delta(\phi)e^{\pi \phi'}}{e^{\pi
\phi'}-e^{\pi \phi}}} d\phi'.$$ The first integral on the
right-hand side is no longer singular, as the integrand approaches
$$
\frac{1}{\pi}\frac{d\delta}{d\phi}
\quad\mbox{as}\quad\phi'\rightarrow\phi.
$$
as $\phi'\rightarrow\phi$.  To see this, note that
$$
\frac{d\delta}{d\phi}=\frac{d\xi}{d\phi}\frac{d\delta}{d\xi}=
\frac{d\xi}{d\phi}
\left[\lim_{\xi'\rightarrow\xi}\frac{\delta(\xi')-\delta(\xi)}{\xi'-\xi}\right]
=\pi
e^{\pi\phi}\left[\lim_{\phi'\rightarrow\phi}\frac{\delta(\phi')-\delta(\phi)}{e^{\pi\phi'}-e^{\pi\phi}}\right].
$$
By using $\phi_N$ as the numerical approximation for infinity
(that is, the greatest value $\phi_I$ in the numerical mesh), the
integral on the right-hand can be integrated directly to give
$$
\int^{\infty}_{0}{\frac{\delta(\phi)e^{\pi \phi'}}{e^{\pi
\phi'}-e^{\pi \phi}}} \, d\phi'\approx \delta(\phi)
\int^{\phi_N}_{0}{\frac{e^{\pi \phi'}}{e^{\pi \phi'}-e^{\pi
\phi}}} \, d\phi' =\frac{\delta(\phi)}{\pi}\ln\left[\frac{e^{\pi
\phi_N}-e^{\pi \phi}}{e^{\pi \phi}-1}\right],
$$
so that we have
\begin{equation}\label{tauff}
\cpvint^{\infty}_{0}{\frac{\delta(\phi')e^{\pi \phi'}}{e^{\pi
\phi'}-e^{\pi \phi}}}
d\phi'\approx\frac{1}{\pi}\int^{\phi_N}_{0}{\frac{\delta(\phi')-\delta(\phi)}{e^{\pi
\phi'}-e^{\pi \phi}}}
d\xi'+\frac{\delta(\phi)}{\pi}\ln\left[\frac{e^{\pi \phi_N}-e^{\pi
\phi}}{e^{\pi \phi}-1}\right].
\end{equation}
It should be noted that the log of zero does not occur as the
above equation is never applied at $\phi_N$.

Now we deal with (\ref{y1}), the equation for relating the known
height of the physical wall to the other flow variables.  By
substituting the known values for $\delta$ from (\ref{delta}),
equation (\ref{y1}) becomes
$$
1-\frac{\mu}{\pi}\int^0_{\xi_B}\frac{e^{-\tau(\xi')}}{\xi'} d \xi'
+\frac{\mu}{\pi}\int^1_{0}\frac{e^{-\tau(\xi')}}{\xi'} d \xi'=0.
$$
Two singularities arise from the limits of the integral in
equation (\ref{y1}) corresponding to stagnation points on the
physical plane located at the origin and directly below the origin
on the channel floor.  With the use of (\ref{tauz}) we have
$$
\frac{e^{-\tau(\xi)}}{\xi}=\frac{1}{(1-\xi)^{1/2}(\xi-\xi_B)^{1/2}}
\,\mbox{exp}\,\left\{-\frac{1}{\pi}\int^{\infty}_{1}{\frac{\delta(\xi')}{\xi'-\xi}}
\,d\xi' \right\}
\quad\mbox{for}\quad \xi_B<\xi<1,
$$
so it's clear that the two stagnation points imply that the
integrand blows up at $\xi=\xi_B$ and $\xi=1$ (but not at
$\xi=0$).  To deal with this singular behaviour we make the
substitution
$$
\xi=\frac{1}{2}[(1-\xi_B)\sin\theta+1+\xi_B],
$$
so that
$$
\frac{d\xi}{(1-\xi)^{1/2}(\xi-\xi_B)^{1/2}}=d\theta,
$$
and hence
\begin{eqnarray*}
\frac{e^{-\tau(\xi)}}{\xi}\,d\xi & = &
\mbox{exp}\,\left\{-\frac{1}{\pi}\int^{\infty}_{1}{\frac{\delta(\xi')}{\xi'-
[(1-\xi_B)\sin\theta+1+\xi_B]/2
}} \,d\xi' \right\}\,d\theta \\
& \approx &
\mbox{exp}\,\left\{-\int^{\phi_N}_{0}{\frac{\delta(\phi')e^{\pi\phi'}}{e^{\pi\phi'}-
\frac{1}{2}[(1-\xi_B)\sin\theta+1+\xi_B]
}} \,d\phi' \right\}\,d\theta \\
& = &
\mbox{exp}\,\left\{-\Lambda(\theta)\right\}\,d\theta,\quad\mbox{say}.
\end{eqnarray*}
The change of variable transforms equation (\ref{y1}) and its
limits of integration accordingly, to the final equation
\begin{equation}\label{y2}
1-\frac{\mu}{\pi}\int_{-\pi/2}^{\theta_B}e^{-\Lambda(\theta')} d
\theta' +
\frac{\mu}{\pi}\int_{\theta_B}^{\pi/2}e^{-\Lambda(\theta')} d
\theta' =0,
\end{equation}
where $\theta_B=-\arcsin((1+\xi_B)/(1-\xi_B))$.  This new equation
has no singularities. However, when numerically calculating
$\Lambda(\frac{\pi}{2})$, a division of zero occurs at $\phi_1=0$.
Fortunately this singularity is removed from the equation as the
exponential term in (\ref{y2}) reduces the integrand at this point
to zero.

\section{Equation solving scheme}
We have $N$ unknowns to  solve with $N$ nonlinear algebraic
equations. In the past the system of equations arising from this
type of problem formulation has been usually solved with an
iterative Newton-Raphson scheme \cite{HF1992, V1996, HF2000}.
Although this method often converges in only a few iterations when
it does find the solution, it requires a \emph{good} initial guess
or otherwise the scheme will not converge. While the
Newton-Raphson scheme is fairly straightforward to implement, each
iteration is quite slow to run in \emph{Matlab} since a $N \times
N$ matrix of linear equations needs to be solved for each
iteration, which is quite computationally exhaustive.

We present alternative solution method by solving the set of $N$
nonlinear equations with the pre-written \emph{Matlab} function
\emph{fsolve}, which uses the Gauss-Newton and Levenberg-Marquardt
schemes depending on the complexity and nature of the system of
equations. As a rough guide the Gauss-Newton scheme is the default
method while the Levenberg-Marquardt scheme is used when the
scheme's step length or condition number goes below a certain
threshold value~\cite{MATLAB}.

Although these two schemes are both more complex than the
Gauss-Newton method, the \emph{fsolve} function is considerably
better at solving nonlinear equations. Compared to the
Newton-Raphson scheme, the initial guess for \emph{fsolve} can be
relatively poor, it takes fewer iterations, and due its
pre-written nature and the use of compiled functions, each
iteration is done in less time.

As a side note, \emph{fsolve} solves a set of homogeneous
equations equations of the form $f_i(x_i)=0$. The final
termination value, determined by the default function tolerance of
\emph{fsolve}, varies for each simulation but for our calculations
is always less than $10^{-15}$, which offers a very reasonable
degree of accuracy for the system.

Thus, we have chosen to solve our system of $N$ nonlinear
equations with the \emph{fsolve} function.

\section{Summary}
We have $N$ unknowns to solve that consist of the $\delta_I$
values for $I=2,3,\ldots,N$, as well as the $\mu$ value. We have
derived $N$ nonlinear equations of which $N-2$ are of the form
$$
\frac{1}{2}\ln\left[\frac{(e^{\pi \phi}-1)(e^{\pi \phi}-e^{\pi
\phi_B})}{e^{2\pi \phi}}\right]
+\cpvint^{\infty}_{0}{\frac{\delta(\phi')e^{\pi\phi}}{e^{\pi\phi'}-e^{\pi\phi'}}}
d\phi'-\frac{1}{3}\ln\left[-\frac{3\mu^3}{
F_{SP}^2}\int_{0}^{\phi}\sin(\delta(\phi'))e^{\pi\phi'}d\phi'\right]=0,
$$
where in order to handle the singularity the integral is expressed
as
$$
\cpvint^{\infty}_{0}{\frac{\delta(\phi')e^{\pi \phi'}}{e^{\pi
\phi'}-e^{\pi \phi}}}
d\phi'\approx\frac{1}{\pi}\int^{\phi_N}_{0}{\frac{\delta(\phi')-\delta(\phi)}{e^{\pi
\phi'}-e^{\pi \phi}}}
d\xi'+\frac{\delta(\phi)}{\pi}\ln\left[\frac{e^{\pi \phi_N}-e^{\pi
\phi}}{e^{\pi \phi}-1}\right].
$$
A further integral equation, which uses the depth of the channel
floor from the stagnation point, was derived as
$$
1-\frac{\mu}{\pi}\int_{-\pi/2}^{\theta_B}e^{-\Lambda(\theta')} d
\theta' +
\frac{\mu}{\pi}\int_{\theta_B}^{\pi/2}e^{-\Lambda(\theta')} d
\theta' =0,
$$
where $\theta_B=-\arcsin((1+\xi_B)/(1-\xi_B))$. The final equation
is a simple extrapolation formula relating the last three $\delta$
values to yield
$$
\delta_N=3\delta_{N-1}-3\delta_{N-2}+\delta_{N-3}.
$$
The final derived system of $N$ equations is solved numerically
using the equation-solving \emph{Matlab} function, \emph{fsolve},
which utilizes Newton-Gauss and Levenberg-Marquardt
schemes~\cite{MORE} and is designed specifically for nonlinear
systems~\cite{MATLAB}.

\chapter{Asymptotic Analysis}\label{chap:asymptotic}
In this chapter we shall obtain an exact expression for $\mu$ when
there are no waves on the free surface, and apply an asymptotic
analysis to the governing integral equation in order to examine
the solution of our problem in the limit $F_{SP}\rightarrow0$.

\section{Upstream height for a solution with no waves}
The purpose of this section is to calculate the average upstream
channel depth when no waves exist upstream on the free surface. In
this case the free surface asymptotes to $y=\mu-1$ as
$x\rightarrow\infty$ with $q\rightarrow\frac{1}{\mu}$ in this
limit.  These values are substituted into the original
nondimensional Bernoulli equation (\ref{fspbernoulli}) to yield
$$
\frac{1}{2}\left(\frac{F_{SP}}{\mu}\right)^2+\mu-1=0,
$$
which can be manipulated to obtain the cubic equation
$$
\mu^3-\mu^2+\frac{1}{2} F_{SP}=0.
$$
This cubic equation is solved using the standard formula (see Abramowitz and
Stegun~\cite{abram}) to yield
$$
\mu=\frac{1}{6}\left(8-54F_{SP}^2+6iF_{SP}\sqrt{24-81F_{SP}^2}
\right)^{\frac{1}{3}}+\frac{2}{3}\left(8-54F_{SP}^2+6iF_{SP}\sqrt{24-81F_{SP}^2}
\right)^{-\frac{1}{3}}+\frac{1}{3},
$$
which can be written more compactly in polar form
\begin{equation}\label{amu}
\mu=\frac{1}{6}r^{1/3}e^{i\theta/3}+\frac{2}{3}r^{-1/3}e^{-i\theta/3}+\frac{1}{3},
\end{equation}
where
\begin{equation}\label{theta}
\theta=\arctan\left(\frac{6F_{SP}\sqrt{24-81F_{SP}^2}}{8-54F_{SP}^2}\right)=
\arccos\left(1-\frac{27}{4}F_{SP}^2\right) ,
\end{equation}
$$
r^2=(8-54F_{SP}^2)^2+36F_{SP}^2(24-81F_{SP}^2)=
64-864F_{SP}^2+2916F_{SP}^2+864F_{SP}^2-2916F_{SP}^2=64.
$$
We use the identity $e^{i\theta} =\cos\theta + i \sin\theta$ to
recast equation (\ref{amu}) to the form
$$
\mu=\left(\frac{1}{6}r^{1/3}+\frac{2}{3}r^{-1/3}\right)\cos\left(\frac{\theta}{3}\right)+
i\left(\frac{1}{6}r^{1/3}-\frac{2}{3}r^{-1/3}\right)\sin\left(\frac{\theta}{3}\right).
$$
Substituting $r=8$ simplifies the expression to
\begin{equation}\label{amuf}
\mu=\frac{1}{3}+\frac{2}{3}\cos\left(\frac{\theta}{3}\right)=
\frac{1}{3}+\frac{2}{3}\cos\left[\frac{1}{3}\arccos\left(1-\frac{27}{4}F_{SP}^2\right)\right].
\end{equation}
Thus, when there are no waves on the free surface, $\mu$ is given
by equation (\ref{amuf}), otherwise $\mu$ is an unknown, which
differs from equation (\ref{amuf}) as the amplitude of the waves
increases.

\section{Regular perturbation method for $F_{SP}\ll 1$}\label{sect:pert}
In this section we shall apply a regular perturbation method to
the nonlinear integral equation (\ref{tauf}) assuming that
$F_{SP}\ll 1$. Hence, we express $\delta(\xi)$ as a power series,
namely
\begin{equation}\label{adelta1}
\delta(\xi)\sim F_{SP}^2\delta_1(\xi)+
F_{SP}^4\delta_2(\xi)+O(F_{SP}^6),
\end{equation}
where $O(F_{SP}^6)$ denotes terms of order $F_{SP}^6$. Thus, we
write the function $\sin\delta(\xi)$ as
\begin{eqnarray}
\sin\delta(\xi) & = & \sin\left[F_{SP}^2\delta_1(\xi)+
F_{SP}^4\delta_2(\xi)+O(F_{SP}^6)\right] \nonumber \\
& \sim & F_{SP}^2\delta_1(\xi)+ F_{SP}^4\delta_2(\xi)+
O(F_{SP}^6). \label{asindelta1}
\end{eqnarray}
We take the exponential of equation (\ref{tauf}) to give
\begin{equation}\label{tauexp}
\left[\frac{(\xi-1)(\xi-\xi_B)}{\xi^2}\right]^{1/2}
\exp\left[\frac{1}{\pi}\cpvint^{\infty}_{1}{\frac{\delta(\xi')}{\xi'-\xi}}
d\xi'\right]=\left[-\frac{3\mu^3}{\pi
F_{SP}^2}\int_{1}^{\xi}\frac{\sin
\delta(\xi')}{\xi'}d\xi'\right]^{1/3}.
\end{equation}
Substituting asymptotic expansions (\ref{adelta1}) -
(\ref{asindelta1}) into the left-hand side of equation
(\ref{tauexp}) yields
\begin{eqnarray*}
& & \hspace{-3ex}
\left[\frac{(\xi-1)(\xi-\xi_B)}{\xi^2}\right]^{1/2}
\exp\left[\frac{F_{SP}^2}{\pi}\cpvint^{\infty}_{1}{\frac{\delta_1(\xi')}{\xi'-\xi}}
d\xi'+O(F_{SP}^4)\right]
\\
& = & \left[\frac{(\xi-1)(\xi-\xi_B)}{\xi^2}\right]^{1/2}\left[1+
\frac{F_{SP}^2}{\pi}\cpvint^{\infty}_{1}{\frac{\delta_1(\xi')}{\xi'-\xi}}
d\xi'+O(F_{SP}^4)\right],
\end{eqnarray*}
while the same substitution into the right-hand side yields, after
using the binomial expansion,
\begin{eqnarray*}
& & \hspace{-3ex} \left[-\frac{3\mu^3}{\pi}\int_{1}^{\xi}\frac{
\delta_1(\xi')}{\xi'}d\xi'-\frac{3\mu^3F_{SP}^2}{\pi
}\int_{1}^{\xi}\frac{
\delta_2(\xi')}{\xi'}d\xi'+O(F_{SP}^4)\right]^{1/3}
\\
& = & \left[-\frac{3\mu^3}{\pi}\int_{1}^{\xi}\frac{
\delta_1(\xi')}{\xi'}d\xi'\right]^{1/3}
\left[1+\frac{\frac{3\mu^3F_{SP}^2}{\pi }\int_{1}^{\xi}\frac{
\delta_2(\xi')}{\xi'}d\xi'}{\frac{3\mu^3}{\pi}\int_{1}^{\xi}\frac{
\delta_1(\xi')}{\xi'}d\xi'}+O(F_{SP}^4)\right]^{1/3}
\\
& = & \left[-\frac{3\mu^3}{\pi}\int_{0}^{\xi}\frac{
\delta_1(\xi')}{\xi'}d\xi'\right]^{1/3}
\left[1+\frac{\frac{\mu^3F_{SP}^2}{\pi }\int_{1}^{\xi}\frac{
\delta_2(\xi')}{\xi'}d\xi'}{\frac{3\mu^3}{\pi}\int_{1}^{\xi}\frac{
\delta_1(\xi')}{\xi'}d\xi'}+O(F_{SP}^4)\right]
\\
& = & \left[-\frac{3\mu^3}{\pi}\int_{1}^{\xi}\frac{
\delta_1(\xi')}{\xi'}d\xi'\right]^{1/3}
+\frac{\frac{\mu^3F_{SP}^2}{\pi }\int_{1}^{\xi}\frac{
\delta_2(\xi')}{\xi'}d\xi'}{\left[\frac{3\mu^3}{\pi}\int_{1}^{\xi}\frac{
\delta_1(\xi')}{\xi'}d\xi'\right]^{2/3}}+O(F_{SP}^4).
\end{eqnarray*}
We take leading powers in the above expansions to obtain
\begin{equation}\label{asym1}
-\frac{3\mu^3}{\pi}\int_{1}^{\xi}\frac{
\delta_1(\xi')}{\xi'}d\xi'=\left[\frac{(\xi-1)(\xi-\xi_B)}{\xi^2}\right]^{3/2},
\end{equation}
and the terms of order $F_{SP}^2$ to give
$$
\frac{\frac{\mu^3}{\pi }\int_{1}^{\xi}\frac{
\delta_2(\xi')}{\xi'}d\xi'}{\left[\frac{3\mu^3}{\pi}\int_{1}^{\xi}\frac{
\delta_1(\xi')}{\xi'}d\xi'\right]^{2/3}}=\left[\frac{(\xi-1)(\xi-\xi_B)}{\xi^2}\right]^{1/2}
\frac{1}{\pi}\cpvint^{\infty}_{1}{\frac{\delta_1(\xi')}{\xi'-\xi}}
d\xi'.
$$
In light of the expression (\ref{asym1}), this last
equation can be rewritten as
\begin{equation}\label{asym2}
-\mu^3\int_{1}^{\xi}\frac{ \delta_2(\xi')}{\xi'}d\xi'
=\left[\frac{(\xi-1)(\xi-\xi_B)}{\xi^2}\right]^{3/2}
\cpvint^{\infty}_{1}{\frac{\delta_1(\xi')}{\xi'-\xi}} d\xi'.
\end{equation}
Thus equation (\ref{asym1}) gives us an equation for the leading
order term $\delta_1$, and with this determined (\ref{asym2}) is
an equation for the correction term $\delta_2$.  In principle, we
could derive equations for higher terms in our expansion
(\ref{adelta1}), however these would soon become increasingly
complex and cumbersome, and impossible to solve analytically.
Thus, we concentrate on solving for $\delta_1$ and $\delta_2$.

To solve for $\delta_1$ we differentiate equation (\ref{asym1}) to
give
$$
-\frac{3\mu^3}{\pi}\frac{
\delta_1(\xi)}{\xi}=\frac{d}{d\xi}\left[\frac{(\xi-1)(\xi-\xi_B)}{\xi^2}\right]^{3/2}=
\frac{3}{2}\frac{\sqrt{(\xi-1)(\xi-\xi_B)}}{\xi^4}(\xi+\xi\xi_B-2\xi_B),
$$
which after rearranging yields
\begin{equation}\label{asymd1a}
\delta_1(\xi)=-\frac{\pi}{2\mu^3}\frac{\sqrt{(\xi-1)(\xi-\xi_B)}}{\xi^3}(\xi+\xi\xi_B-2\xi_B).
\end{equation}
This expression was derived by Hocking and Forbes~\cite{HF1992}.

We now extend the result of Hocking and Forbes~\cite{HF1992} by solving for $\delta_2$.
To do this we need to substitute the expression for
$\delta_1$ (\ref{asymd1a}) into (\ref{asym2}), which involves
calculating
$$\cpvint_1^{\infty}{\frac{\delta_1(\xi')}{\xi'-\xi}} d\xi'=
-\frac{\pi}{2\mu^3}\cpvint_1^{\infty}{\frac{\sqrt{(\xi'-1)(\xi'-\xi_B)}
(\xi'+\xi'\xi_B-2\xi_B)}{\xi'^3(\xi'-\xi)}}\,d\xi'.
$$
We calculate the above integral using the computer algebra
manipulation package \emph{Maple} to yield
\begin{equation}\label{intasym}
\cpvint_1^{\infty}{\frac{\sqrt{(\xi'-1)(\xi'-\xi_B)}
(\xi'+\xi'\xi_B-2\xi_B)}{\xi'^3(\xi'-\xi)}}\,d\xi'=\left[U\arctan(X)+W\ln\left|Y\right|-Z\right]_1^{\infty},
\end{equation}
which has been simplified by introducing the five terms
\begin{equation}\label{U}
U=-\frac{1}{4}\frac{1}{\xi^3\sqrt{(-\xi_B)}}
\left(\xi^2+\xi^2\xi_B^2-8\xi\xi_B^2-8\xi\xi_B+6\xi^2\xi_B+8\xi_B^3\right),
\end{equation}
\begin{equation}\label{W}
W=-\frac{1}{\xi}\sqrt{(\xi-1)(\xi-\xi_B)} (\xi+\xi\xi_B-2\xi_B),
\end{equation}
\begin{equation}\label{X}
X=\frac{\xi'+\xi'\xi_B-2\xi_B}{2\sqrt{(-\xi_B)}\sqrt{(\xi'-1)(\xi'-\xi_B)}},
\end{equation}
\begin{equation}\label{Y}
Y=\frac{-\xi'-\xi'\xi_B+2\xi_B-\xi'-\xi'}{\xi'-\xi},
\end{equation}
\begin{equation}\label{Z}
Z=-\frac{1}{2}\frac{1}{\xi^2\xi'^2}\left(2\xi\xi_B-3\xi\xi_B\xi'-3\xi\xi'+4\xi'\xi_B\right)\sqrt{(\xi'-1)(\xi'-\xi_B)}.
\end{equation}
The limits are applied to the above expression to starting with
(\ref{X}), which yields
$$
X\rightarrow\infty \quad \textrm{as}\quad \xi'\rightarrow1^+,\quad
\textrm{thus,}\quad \arctan(X)\rightarrow\frac{\pi}{2},
$$
$$
X\rightarrow\frac{1+\xi_B}{2\sqrt{-\xi_B}}\quad \textrm{as}\quad
\xi'\rightarrow\infty,\quad \textrm{thus,}\quad
\arctan(X)\rightarrow\arctan\left(\frac{1+\xi_B}{2\sqrt{-\xi_B}}\right)=\frac{\pi}{2}-2\arctan\sqrt{-\xi_B},
$$
where the last result was rewritten with a simple trig identity. The limits are applied to equations (\ref{Y})-(\ref{Z})
give
$$
Y\rightarrow-1+\xi_B \quad \textrm{as}\quad
\xi'\rightarrow1^+,\quad \textrm{thus,}\quad
\ln|Y|\rightarrow\ln\left|1-\xi_B \right|,
$$
$$
Y\rightarrow-1-\xi_B+2\xi+2\sqrt{(\xi-1)(\xi-\xi_B)}\quad
\textrm{as}\quad \xi'\rightarrow\infty,
$$
$$
Z\rightarrow0 \quad \textrm{as}\quad \xi'\rightarrow1^+,
$$
$$
Z\rightarrow\frac{3}{2}\frac{\xi_B}{\xi}+\frac{3}{2\xi}-\frac{2\xi_B}{\xi^2}
\quad \textrm{as}\quad \xi'\rightarrow\infty.
$$
These results are substituted into expression (\ref{intasym}) to
yield
$$
\cpvint_1^{\infty}{\frac{\delta_1(\xi')}{\xi'-\xi}}
d\xi'=\frac{1}{2}\frac{1}{\xi^3\sqrt{(-\xi_B)}}
\left(\xi^2+\xi^2\xi_B^2-8\xi\xi_B^2-8\xi\xi_B+6\xi^2\xi_B+8\xi_B^3\right)\arctan\sqrt{-\xi_B}...
$$
\begin{equation}\label{intasym2}
-\frac{1}{\xi}\sqrt{(\xi-1)(\xi-\xi_B)}
(\xi+\xi\xi_B-2\xi_B)\ln\left|\frac{-1-\xi_B+2\xi+2\sqrt{(\xi-1)(\xi-\xi_B)}}{1-\xi_B}\right|
-\frac{3}{2}\frac{\xi_B}{\xi}+\frac{3}{2\xi}-\frac{2\xi_B}{\xi^2}.
\end{equation}
Thus, we  obtained an expression for $\delta_2$ by substituting
(\ref{intasym2}) into equation (\ref{asym2}), and differentiating
the result in \emph{Maple} to give
\begin{eqnarray}
\delta_2(\xi) & = & -\frac{\xi}{\mu^3}\frac{d}{d\xi}\left(\left[\frac{(\xi-1)(\xi-\xi_B)}{\xi^2}\right]^{3/2}
\cpvint^{\infty}_{1}{\frac{\delta_1(\xi')}{\xi'-\xi}} d\xi'\right)
\nonumber \\
& = & \frac{\pi}{8\mu^6}
\left.\left.\frac{\sqrt{(\xi-1)(\xi-\xi_B)}}{\xi^5}
\right\{
\right[(2\xi_B^2+12\xi_B+2)\xi^4-(\xi_B+1)(5\xi_B^2+62\xi_B+5)\xi^3
\nonumber \\
& & +16\xi_B(4\xi_B^2+
13\xi_B+4)\xi^2-152\xi_B^2(\xi_B+1)\xi+96\xi_B^2\left]
\frac{\arctan\sqrt{-\xi_B}}{\xi\sqrt{-\xi_B}}\right.
\nonumber \\
& & + 10(1+\xi_B)\xi^3-(19\xi_B^2+62\xi_B+19)\xi^2+64\xi_B(\xi_B+1)-48\xi_B^2
\nonumber \\
& &
\left.\left.-\frac{4\sqrt{(\xi-\xi_B)(\xi-1)}}{\xi}\right[(\xi_B+1)\xi^3-(3+\xi_B)(3\xi_B+1)\xi^2+13\xi_B(\xi_B+1)\xi
-12\xi_B^2\right]
\nonumber \\
& & \left.\times
\log\left(\frac{2\xi-1-\xi_B+2\sqrt{(\xi-1)(\xi-\xi_B)}}{1-\xi_B}\right)\right\}.
\label{asymd2a}
\end{eqnarray}
Note that if the source is situated on on the channel floor then $\xi_B=0$.  By taking the
limit $\xi_B\rightarrow 0^-$ above we find that
(\ref{asymd1a}) and (\ref{asymd2a}) reduce to the rather simple expressions
\begin{equation}\label{asymd1b}
\delta_1(\xi)=-\frac{\pi}{2\mu^3}\frac{\sqrt{(\xi-1)\xi}}{\xi^2},
\end{equation}
\begin{equation}\label{asymd2b}
\delta_2(\xi)=\frac{\pi}{2\mu^6}\left(\frac{3\sqrt{\xi-1}}{\xi^{5/2}}(\xi-2)-\frac{(\xi-1)(\xi-3)}{\xi^3}
\ln\left[2\xi-1+2\sqrt{\xi(\xi-1)}\right]\right).
\end{equation}
These expressions are substituted into the original power series
(\ref{adelta1}) to a asymptotic expression for $\delta(\xi)$ that
serves as a good initial guess to our numerical scheme.

\section{Asymptotic description which includes waves}
The regular perturbation method outlined in section
\ref{sect:pert} predicts a horizontal free surface in the far
stream. However, numerical results in chapter \ref{sect:numasym}
(see figure \ref{xyfsp045}) suggest that a train of waves does
exist on the free surface, whose amplitude is exponentially small
as $F_{SP}$ approaches zero. This would explain why our regular
perturbation method does not capture the waves since
 $\exp(-\beta\mbox{/} F_{SP}^2)$ cannot be written as a power series
in $F_{SP}^2$~\cite{B1999}. Hence, this implies that the free
surface waves are \emph{beyond all orders}, or equivalently, they
are smaller than every term in the regular perturbation series
\begin{equation}\label{adelta2}
\delta(\xi)= \sum_{n=1}^{\infty} F_{SP}^{2n}\delta_n(\xi).
\end{equation}

Thus, in order to describe the free surface completely,
including waves, in the limit $F_{SP}\rightarrow 0$, we must employ
\emph{exponential asymptotics}. Unfortunately, this is a highly
complex and rigorous theory that is beyond the scope of an honours
project. In lieu of this we briefly outline how this method could
be employed.

Firstly, it should be noted that in principle it is possible to
derive any number of terms in a series like (\ref{adelta2}), but
in practice any more than two terms would become nearly
impossible, even if employing symbolic packages like \emph{Maple}.
Typically, for a fixed value of $F_{SP}$ such a series would be divergent
\cite{B1999}.  Furthermore, the error will likely decrease as more terms are included up to a certain point,
but then increase without bound.  Thus, it should be
possible to optimally truncate the series in order to minimized
the error. This optimally truncated asymptotic series is sometimes
referred to as \emph{superasymptotic}~\cite{B1999}.

The crucial point here is that the error in a superasymptotic
series is typically of the order $\exp(-\beta\mbox{/} \epsilon)$,
where $\epsilon$ is a small parameter (in our case
$\epsilon=F_{SP}^2$). These exponentially small terms form part of
an \emph{subdominant exponential}, and are hidden from normal
view, but are ``switched on'' across a Stokes' line. This
behaviour is closely related to Stokes' phenomena, which states
that a single valued analytic function may have a different
asymptotic expansion in different sections of the complex
plane~\cite{Thomson}.

We mention that exponentially small capillary waves (due to
surface tension but not gravity) have been successfully described
using exponential asymptotics by Chapman and Vanden-Broeck
\cite{CV2002} by first deriving a Nekrasov type integral equation
(analogous to equation (\ref{tauf})). However, our type of problem
has not been studied with asymptotic exponentials hence this
remains a goal for future research.

\chapter{Presentation of Results}\label{chap:results}
In this chapter we shall present a number of typical numerical and
asymptotic results for varying $F_{SP}$ values and discuss their
solution characteristics. We shall examine a numerical solution
with waves and make a comparison between different values of $N$
and $\phi_N$ in order to argue that our chosen parameter values
are suitable for our calculations. We shall compare the numerical
and asymptotic solutions and outline the $F_{SP}$ range where they
agree and where the asymptotic solution begins to break down.
Finally, we shall analyze the behaviour of wave amplitude and
wavelength for small values of $F_B$, recalling that $F_B$ is
related to $F_{SP}$ by the expression
$$
F_B=\frac{F_{SP}}{\mu^{3/3}}.
$$
For an example, we have chosen to calculate solutions for a source
located on the channel floor which corresponds to $\xi_B=0$.
Interestingly, this is also equivalent to two-dimensional problem
of flow past a ship stern (see section \ref{subsect:stern}) for
the limit $H\rightarrow 0$.

\begin{figure}[h]
\begin{center}
\includegraphics[height=8.0cm,width=10cm]{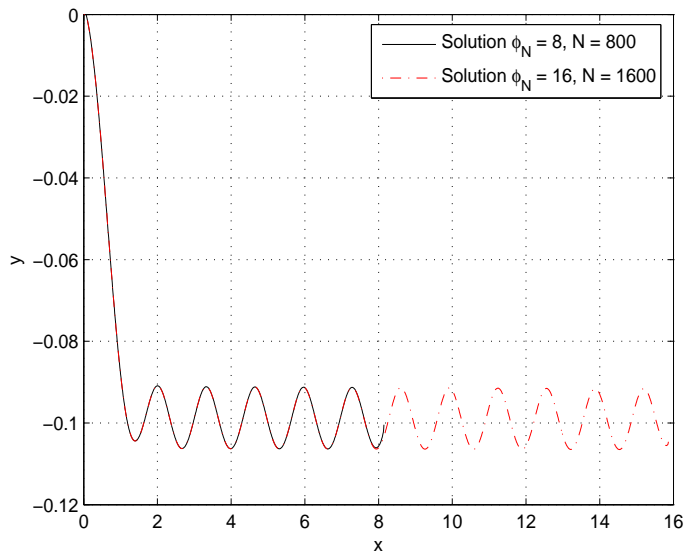}
\caption{A comparison of free surface profiles to observe the
effect of different approximations for infinity. $F_{SP}=0.4$.
\label{phicomp}}
\end{center}
\end{figure}

\section{Effects of numerical scheme parameters}
\begin{figure}[h]
\begin{center}
\includegraphics[height=8.0cm,width=10cm]{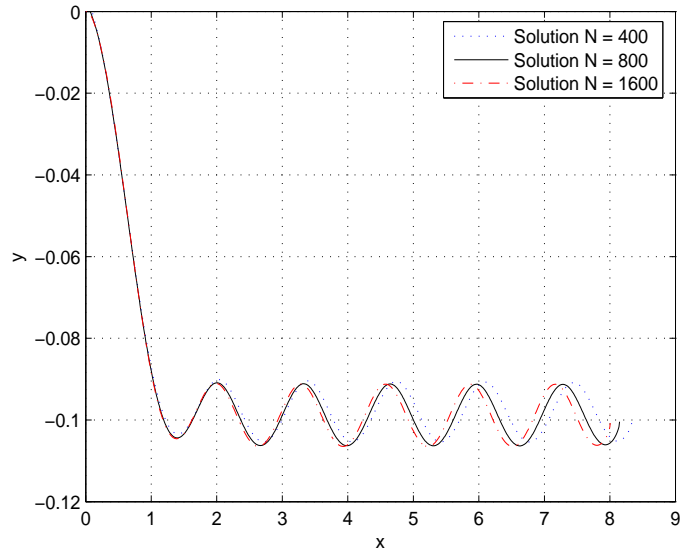}
\caption{The effect on the numerical solution when the number of
mesh points is varied. $F_{SP}=0.4$, $F_B=0.4677$ and
$\mu=0.9009$. \label{ncomp}}
\end{center}
\end{figure}

We wish to use the most appropriate $\phi_N$ and $N$ parameter
values in order to gain a solution with a reasonable degree of
numerical accuracy. We have chosen $\phi_N=8$ and $N=800$, and
presented a typical solution $F_{SP}=0.4$ in figure \ref{ncomp},
which shows a train of waves present upstream on the free surface.
In order to justify this approximation for infinity, we present a
second solution of $F_{SP}=0.4$ with $\phi_N=16$. In an attempt to
counter the truncation errors of the coarser mesh for the
$\phi_N=16$ solution, the number of points is doubled to yield
$N=1600$, which ensures that the $\Delta\phi$ is the same value
for both solutions.

The plot shows the two solutions agreeing remarkably well. The
$\phi_N=8$ solution is almost identical to the $\phi_N=16$
solution. Seemingly, the larger $\phi_N$ value has no observable
advantage and thus, we shall assume that $\phi_N=8$ is a good
approximation for infinity in our system.

Next we compare the solutions for $F_{SP}=0.4$ when $N=400$,
$N=800$ and $N=1600$ in order to see what effect increasing the
number of mesh points has on the solution. The three solutions
agree with each other for low values of $x$, particularly before
the first trough. However, the solutions begin to differ with an
increase in $x$. Naturally, the $N=400$ solution differs more than
the $N=800$ solution in comparison to the $N=1600$ solution. It is
assumed that $N=1600$ solution is more accurate. However, it
should be noted that when too many mesh points are used round-off
errors become significant, thus reducing the overall accuracy of
the system~\cite{NUMR}. Although, it is not certain in which
region of $N$ this problem manifests itself.

An important consideration, other than numerical accuracy, is the
pragmatic concern of computation time. To give a rough measurement
of computation time, the majority of these calculations were
performed on a $1.3$ GHz machine, which calculated the solution of
a typical system ($F_{SP}=0.35$, $\phi_N=8$ and $N=800$) in $113$
minutes with an average of five to six iterations.

Intuitively, when the mesh becomes finer the numerical scheme
produces more accurate results. This advantageous increase in
accuracy is countered by the obvious increase in the number of
equations that need to be solved. Thus, the task of choosing the
$N$ value, becomes a trade off between accuracy and computation
time.

In conclusion, we found that $\phi_N=8$ is a good approximation
for infinity while $N=800$ is very reasonable number of points and
that increasing this value did not have a significant advantage
when computation time was taken under consideration. Henceforth,
we shall use the parameters $\phi_N=8$ and $N=800$ in all the
ensuing solutions presented here unless stated otherwise.

\section{Comparison of numerical and asymptotic solutions}\label{sect:numasym}
Our asymptotic scheme, which we derived in
chapter~\ref{chap:asymptotic}, is limited to the regime of low
values of $F_{SP}$ and is incapable of producing solutions which
possess waves on the free surface. We present a comparison of the
numerical and asymptotic solutions.
\begin{figure}[h]
\begin{center}
\includegraphics[height=7.0cm,width=9cm]{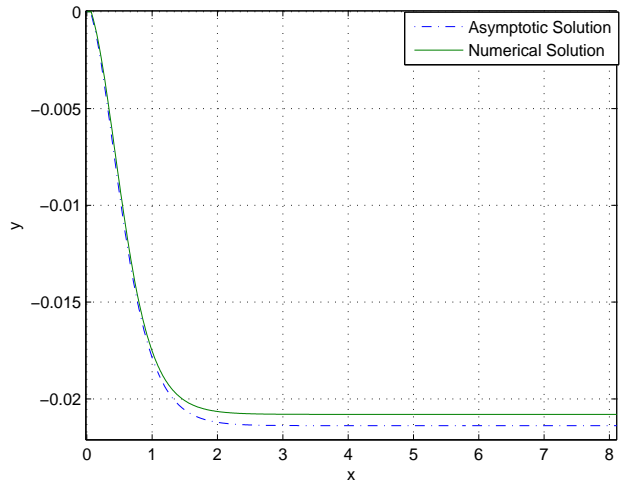}
\caption{Comparison of asymptotic and numerical free surface
profiles for $F_{SP}=0.2$.\label{xyfsp02}}
\end{center}
\end{figure}

\begin{figure}[h]
\begin{center}
\includegraphics[height=7.0cm,width=9cm]{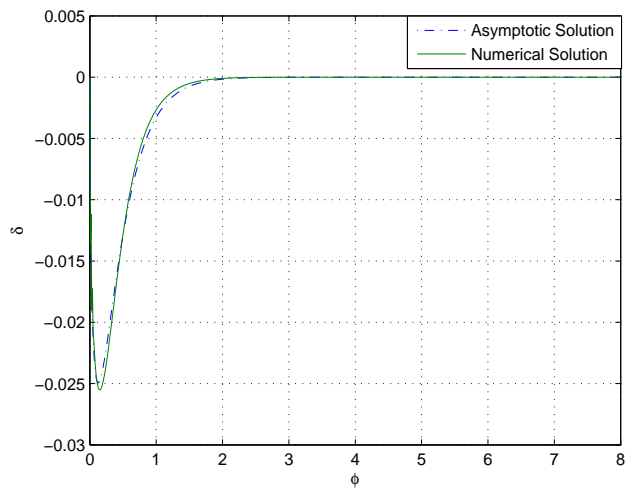}
\caption{Comparison of asymptotic and numerical $\delta$ solutions
for $F_{SP}=0.2$.\label{deltafsp02}}
\end{center}
\end{figure}

The plots of the free surface profile and the angle of the free
surface, $\delta$ in figures \ref{xyfsp02} - \ref{deltafsp02} show
that the asymptotic and numerical solutions are in close agreement
for $F_{SP}=0.2$. Although there are exponentially small waves
present on the free surface of the numerical solution, which we
discuss in the next section, the profile is virtually flat for
this $F_{SP}$ value. Hence, the asymptotic approach serves as a
good solution and an excellent initial guess for the numerical
solution.

\begin{figure}[h]
\begin{center}
\includegraphics[height=7.0cm,width=9cm]{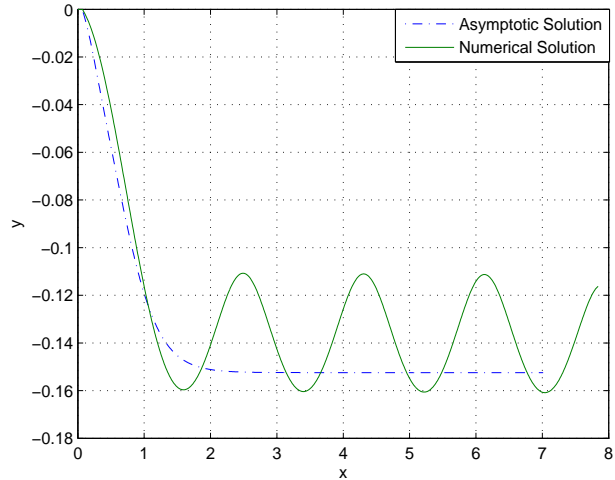}
\caption{The asymptotic solutions breaks down for $F_{SP}=0.45$.
\label{xyfsp045}}
\end{center}
\end{figure}

In figure \ref{xyfsp045} a similar plot of the asymptotic and
numerical solutions for $F_{SP}=0.45$ shows that they are not in
agreement, which is to be expected as the asymptotic scheme
becomes more inaccurate as $F_{SP}$ increases. In particular,
there are sizeable waves present on the free surface while, as
previously noted, our asymptotic solution is unable to produce a
solution with waves.

Near the stagnation point, the plots of asymptotic and numerical
$\delta$ agree with each other. However, as the asymptotic
$\delta$ approaches zero in the far field (the free surface
plateaus), the numerical $\delta$ continues to oscillate above and
below zero corresponding to sizeable waves on the free surface.
The asymptotic solution offers a reasonable $\delta$ guess for
smaller waves. However, guesses such as $\delta$ solutions for
smaller $F_{SP}$, should be considered instead of asymptotic
$\delta$ solutions when dealing with larger $F_{SP}$ values.
\begin{figure}[h]
\begin{center}
\includegraphics[height=7.0cm,width=9cm]{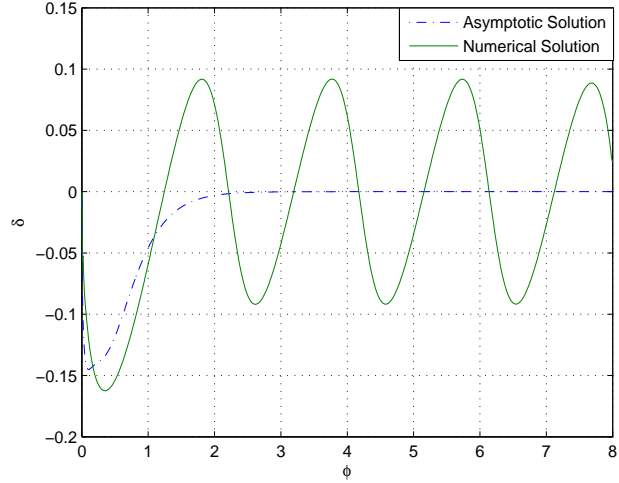}
\caption{Asymptotic solution is unable to predict waves for
$F_{SP}=0.45$\label{deltafsp045}}
\end{center}
\end{figure}

\section{Wave behaviour}
We wish to shed light on the behaviour of the train of waves
present on the free surface profile. In particular, we wish to
analyse the variation of wave amplitude, $A$ and wavelength
$\lambda$ with respect to $F_B$.

\begin{figure}[h]
\begin{center}
\includegraphics[height=7.0cm,width=9cm]{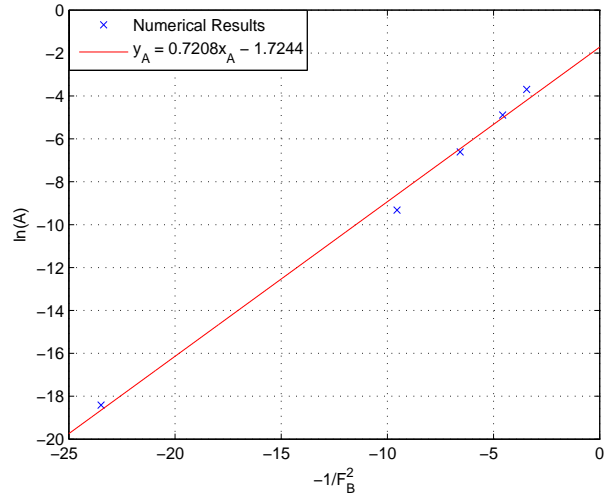}
\caption{A linear fit of the numerical results that suggest $A$ decreases
exponentially. \label{fbamp}}
\end{center}
\end{figure}

Mekias and Vanden-Broeck  in their original paper \cite{MV1991}
observed for small values of $F_B$ an apparent linear relationship
existing between $-1/F_B^2$ and  $\ln(A)$. By plotting $\ln(A)$
versus $-1/F_B^2$, we observed a similar linear relationship in
figure \ref{fbamp}, which suggests
\begin{equation}\label{amprelation}
A=\alpha e^{-\beta/F_B^2} \quad \textrm{as}\quad F_B\rightarrow0,
\end{equation}
where $\alpha$ and $\beta$ are constants of the individual system
and are, according to Mekias and Vanden-Broeck~\cite{MV1991},
related to the location of the source.

Since the amplitude of the free surface waves decreases rapidly
due to their exponential nature, waves are barely visible for
smaller values of $F_B$, and hence the profiles appear flat.
Vanden-Broeck suggests in his reconciliation work~\cite{V1996}
that this may be the reason why waves were not visible in the
results of Hocking and Forbes~\cite{HF1992}. We tend to agree with
this conclusion since our results show seemingly flat profiles for
values such as $F_B=0.2064$.
\begin{figure}[h]
\begin{center}
\includegraphics[height=7.0cm,width=9cm]{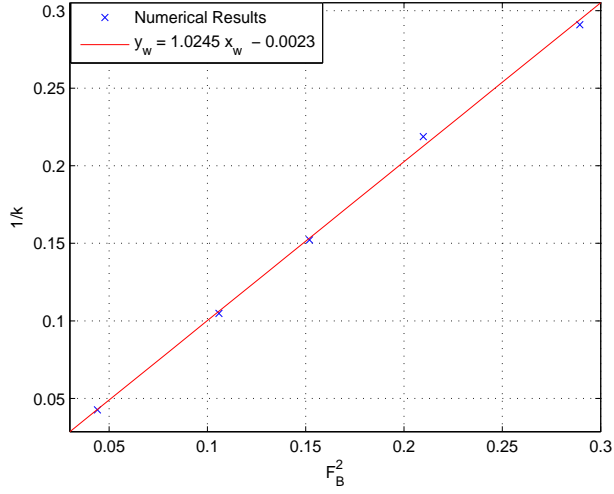}
\caption{A fitted line with an almost unity gradient suggesting an exact
expression for $k$.\label{fblambda}}
\end{center}
\end{figure}

Our results demonstrate that free surface waves exhibit a periodic
behaviour. In order to illuminate this periodic nature, we
introduce the wave number, defined as $k=2\pi/\lambda$. For the
regime of $0.2 \leq F_{SP}\leq 0.45$, we plot $1/k$ versus $F_B^2$
and obtain an elegant linear relationship. The corresponding
fitted-line has a gradient of almost unity, which strongly
suggests
\begin{equation}\label{lrelation}
\frac{1}{k}=\frac{\lambda}{2\pi}=F_B^2 \quad \textrm{as}\quad
F_B\rightarrow0.
\end{equation}
The equivalent relationship based on linear wave theory is
\begin{equation}\label{lrelation2}
F_B^2=\frac{1}{k}\tanh(k),
\end{equation}
which reduces to our empirical expression (\ref{lrelation}) as we
approach the limit
$$
\tanh(k)\rightarrow 1\quad \textrm{as}\quad \frac{1}{k}\rightarrow
0.$$ Thus, this relationship is a reliable vindication of our
results and overall solution method.
\begin{figure}[h]
\begin{center}
\includegraphics[height=7.0cm,width=9cm]{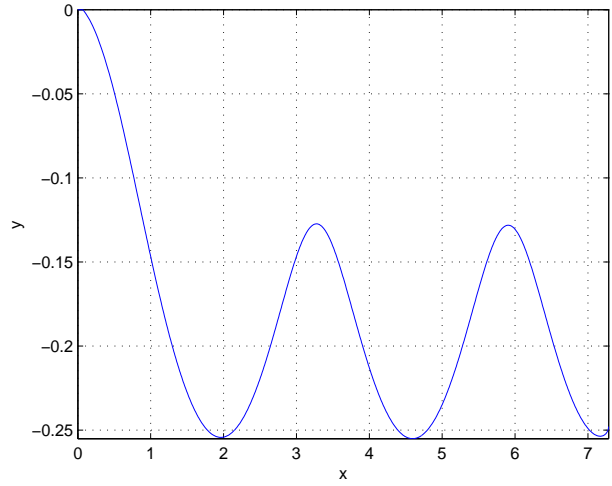}
\caption{Narrower crests and broader troughs for $F_{SP}=0.5$. \label{xyfsp05}}
\end{center}
\end{figure}

\begin{figure}[h]
\begin{center}
\includegraphics[height=7.0cm,width=9cm]{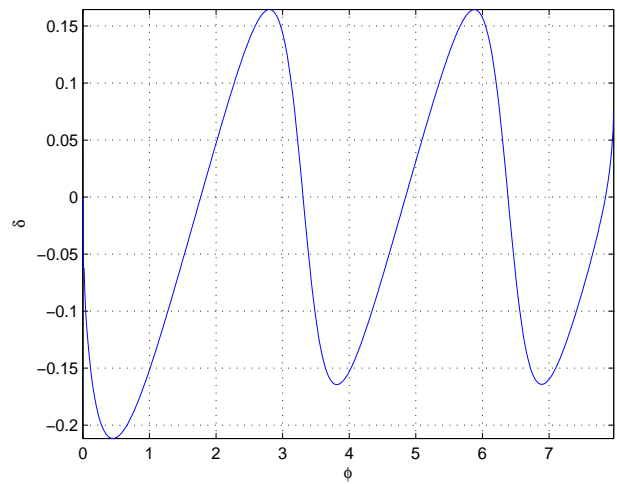}
\caption{The
free surface angle, $\delta$ becoming skewed for
$F_{SP}=0.5$.\label{deltafsp05}}
\end{center}
\end{figure}

Finally, our results in figures (\ref{xyfsp05}) and
(\ref{deltafsp05}) reveal that the general form of the waves
continues to change as $F_{SP}$ increases. The waves are no longer
sinusoidal in shape, but begin to have narrow crests and broad
troughs. For larger Froude numbers the waves will approach the
Stokes' limiting configuration with $120$-degree
crests~\cite{RAHMAN}. However, the number of points needed for
these types of calculations is more than feasible.

\chapter{Conclusion and Discussion}\label{chap:conclude}

\section{Summary of project}
In this project we have presented a lucid formulation of a free
surface problem with flow due to a submerged source in a channel
of finite depth. We have clearly declared all our assumptions and,
correspondingly, derived all the necessary governing equation and
boundary conditions. In the process, we have rectified and
clarified any errors and misconceptions.

We have employed complex variable theory in a similar manner as
others have done before us to solve free surface
problems~\cite{HF1992, V1996, HF2000}. Using conformal mapping and
Cauchy's integral theorem, we have obtain a singular nonlinear
integral equation. This integral equation has been solved by
representing it with trapezoidal scheme and solving the resulting
set of nonlinear equations with the \emph{fsolve} function and in
the process we have carefully dealt with all the singularities .

We have extended the work of Hocking and Forbes \cite{HF1992} by
deriving a second order asymptotic solution for $\delta(\xi)$,
which is valid for the regime of small $F_B$, and as such, the
asymptotic results have agreed well with the numerical results in
this regime. We have discussed how exponential asymptotics could
be used to derive a more accurate analytic solution that describes
waves on the fee surface.

We have obtained results showing waves of significant amplitude on
the free surface, which is in agreement of previous
solutions~\cite{HF1992, V1996,HF2000}. We have studied the
behaviour of the free surface waves in relation to the $F_B$ value
and have verified and obtained a number of relationships, which
agree with previous observations~\cite{V1996} and linear wave
theory~\cite{RAHMAN}. Thus, we have vindicated our results and
solution methods with these empirical observations and
relationships.

\section{Related free surface problems}\label{sec:relatedfsproblems}
It happens that the problem of flow due to a line source is
closely related to a number of other free surface problems, such
as the flow past the stern of a ship, and the flow under a sluice
gate. As a result, the solution methods employed here may be
adapted for these related problems. We briefly discuss the
problems here.

\subsection{Flow past the stern of a ship}\label{subsect:stern}
Evidently, when ships travel through water they create waves. This
wave creation is generally unwanted as it creates drag, which
results in more energy being spent to propel the ship. For naval
vessels, waves are unwanted for an additional reason as waves
leave a ``foot print'' thus allowing it to be detected by an
enemy. Thus, ships should be designed in order to minimize wave
generation and hence drag.
\begin{figure}[!htb]
\begin{center}
\includegraphics[height=5.0cm,width=9cm]{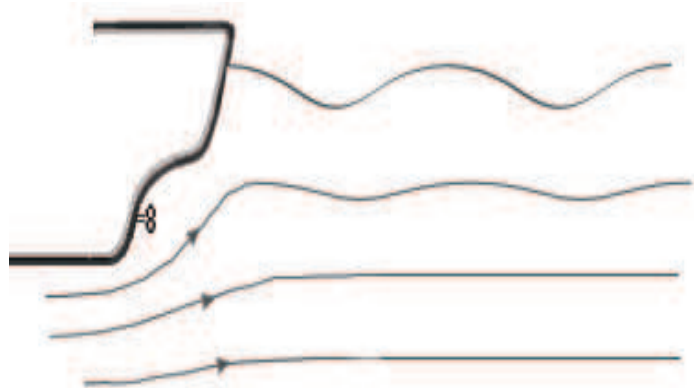}
\caption{Flow past a regular ship stern.\label{ship}}
\end{center}
\end{figure}

In two dimensions we can study the flow past a stern of a ship by
assuming that the stern is  semi-infinite, as in figure 7 of
Tuck's~\cite{T1991} (reproduced here as figure \ref{ship}).
Numerous problems of this nature have been analyzed
numerically~\cite{T1991,M1988,FT1995,MF1999,VT1977}. For example,
various stern shapes have been tried in order to minimize the
amplitude of the waves they create, with varying success. These
studies have been numerical in nature, with no precise understanding
of the exact relationship between wave amplitude and stern shape.
\begin{figure}[!htb]
\begin{center}
\includegraphics[height=6.5cm,width=10cm]{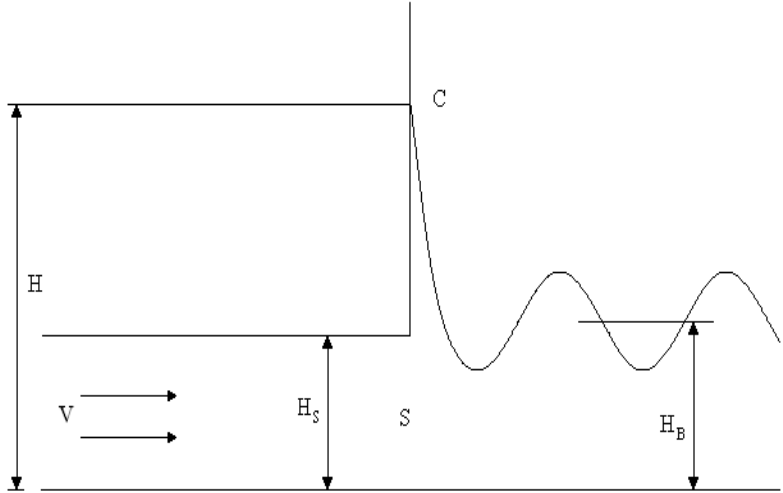}
\caption{Flow past
a rectangular ship stern.\label{stern}}
\end{center}
\end{figure}

In order to see how similar the problem presented in this thesis
is we consider the simple case when the ship stern is rectangular
in shape as indicated in figure \ref{stern}. We scale the
variables to yield
$$x=\frac{\bar{x}}{H}, \quad y=\frac{\bar{y}}{H},\quad\mu=\frac{H_B}{H},
\quad\gamma=\frac{H_S}{H},$$ and scale the velocities with respect
to $V\lambda$, so
$$
q\rightarrow\frac{1}{\lambda}\quad \textrm{as}\quad
x\rightarrow-\infty.
$$
The Bernoulli equation, which is satisfied anywhere along the free
surface, becomes
$$
\frac{1}{2}\left(\frac{m^2}{gH^3}\right)q^2+y=0.
$$
We reintroduce the complex potential $f(z)=\phi(x,y)+i\psi(x,y)$,
and expressed the complex velocity as
$$
\frac{df}{dz}=w=u-iv.
$$
The problem is mapped to the $f$-plane, then to the $\zeta$-plane.
The complex velocity is now expressed as
$$
w=\frac{1}{\mu}e^{-i\Omega(\zeta)},
$$
where $\Omega(\zeta) = \delta(\zeta) + i\tau(\zeta)$. The flow
conditions of the physical problem give the flow angles
$$\delta (\xi) = \left\{ \begin{array}{ccc}
  0 & \textmd{if} & -\infty < \xi < \xi_S, \\
  \pi/2 & \textmd{if} & \xi_S <\xi < 1, \\
  \textmd{unknown} & \textmd{if} &  \xi >1.
\end{array}\right.
$$
The Bernoulli equation is recast and manipulated to become
$$
\tau(\xi)=\frac{1}{3}\ln\left[-\frac{3\mu^3}{\pi
F_{SP}^2}\int_1^{\xi}\frac{\sin \delta(\xi')}{\xi'}d\xi'\right].
$$
We invoke the Cauchy integral formula and use the aforementioned
$\delta(\xi)$ values to obtain the integral equation
$$
\tau(\xi)=\frac{1}{2}\ln\frac{(\xi-1)}{(\xi-\xi_B)}
+\frac{1}{\pi}\cpvint^{\infty}_{1}{\frac{\delta(\xi')}{\xi'-\xi}}
d\xi',
$$
which is combined with the Bernoulli equation to yield
$$
\frac{1}{2}\ln\frac{(\xi-1)}{(\xi-\xi_B)}
+\frac{1}{\pi}\cpvint^{\infty}_{1}{\frac{\delta(\xi')}{\xi'-\xi}}
d\xi' -\frac{1}{3}\ln\left[-\frac{3\mu^3}{\pi
F_{SP}^2}\int_1^{\xi}\frac{\sin \delta(\xi')}{\xi'}d\xi'\right]=0.
$$
This integral equation is almost identical to our integral
equation (\ref{tauf}), and as such, can be solved with an approach
similar to ours.
\begin{figure}[h]
\begin{center}
\includegraphics[height=5.0cm,width=8cm]{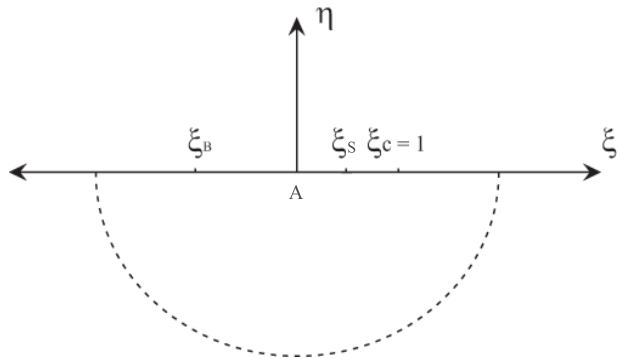}
\caption{The lower half of the $\zeta$-plane. \label{zetastern}}
\end{center}
\end{figure}

\subsection{Flow under a sluice gate}
\begin{figure}[!htb]
\begin{center}
\includegraphics[height=6.5cm,width=11cm]{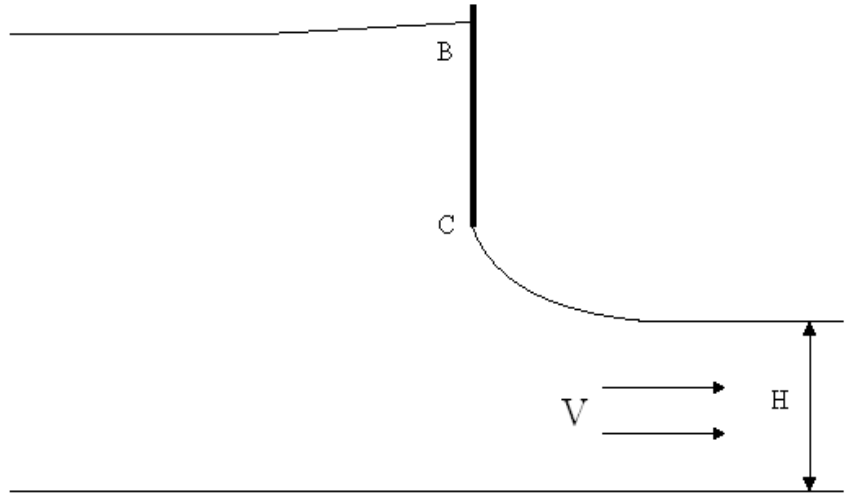}
\caption{Flow under a sluice gate with two free surfaces.\label{sluice}}
\end{center}
\end{figure}

The flow under a sluice gate is a classical problem in applied
mathematics, studied by \cite{V1997, FS1968}, and many others.
This problem is similar to the flow due to a source and flow past
a stern because of the stagnation point (see point B on figure
\ref{sluice}). This problem is more difficult to solve due to the
presence of the two free surfaces, however, conformal mapping can
be used similarly to give an integral equation analogous to ours.
We omit the details, but instead refer the reader to \cite{V1997}.

\section{Further research}
\subsection{Exponential Asymptotics}
Since our asymptotic analysis is apparently incapable of capturing
the exponentially small terms that produce the free surface waves,
it seems the terms are beyond all orders. However, it may be
possible to shed more light on on this area by employing
exponential asymptotics in order to derive an asymptotic expansion
that captures the free surface waves for small Froude numbers.
This would be indeed an exciting and difficult task for future
research.

\subsection{Flow due to a sink}
If it is possible to show with exponential asymptotics that a train
of waves, albeit exponentially small, always exist on the free
surface then, by the Sommerfield radiation condition, we can infer
that the flow field is due to a line source, and cannot be used to describe flow due to a line sink.
Thus, the question remains, what is the flow configuration if there is a
link sink, and not a line source, present? As mentioned in section
\ref{sect:flow}, this is an important question in regards to fluid
withdrawal of various reservoirs. In order to answer question,
further research is required, particularly for the case of Froude
numbers less than unity, as this problem remains open

\subsection{Stern flows}
We have illustrated in section \ref{subsect:stern} how our problem
is very similar to that of flow past a ship stern in two
dimensions. Hence, we expect for this case that exponentially
small waves would exist on the free surface for small Froude
numbers. Similarly, these waves could possibly be described with
the use of exponential asymptotics although this would be an
arduous task. However, an exciting prospect of employing
exponential asymptotics is the possibility of deriving analytical
results that relate the shape of stern and the amplitude of the
waves with the goal of designing an optimally shaped stern. This
problem, along with the others discussed, remains open for future
endeavours.

\newpage
\appendix

\chapter{Numerical code}\label{chap:code}
\section{\emph{froudesolve.m}}
\begin{verbatim}
%Main Program
%Makes asymptotic initial guess, solves equations for delta (hock.m),
%calculates tau, calculates x and y for free surface, plots results
clear all; close all; t1=cputime; load nasymphi8Fs04; Fsp=0.4;
zb=0; n=100;
%Setting up the zeta values
phimax=12; phi=linspace(0,phimax,n); z=exp(pi*phi); zmax=z(end);
%Initial Guess
mu=1/3+2/3*cos(1/3*acos(1-27/4*Fsp^2));
guess=Fsp^2*(-pi/2/mu^3).*sqrt((z-1).*(z-zb))./z.^3.*(z+zb*(z-2));
guess(1)=mu;%mu guess

%fslove
options=optimset('Display','iter');   % Option to display output
%Minimising equations
guess= fsolve(@froude,guess,options); %Make sure the parameters match
%Retrieve calculated values
mu=guess(1); ndelta=[0 guess(2:end)];

%Calculate tau values
tau=[]; const=-3*mu^3/Fsp^2; for j=2:n
    tau(j)=(1/3)*log((const*trapezoid(phi(1:j),sin(ndelta(1:j)))));
end
tau=[0 tau]; %Zero is for adjusting length
%Calculate x and y
x=[];y=[];x(1)=0;y(1)=0;y(2)=0; j=2;
%Integrand before the singularity
fL=(ndelta(1:j-1)-ndelta(j))./(exp(pi*phi(1:j-1))-exp(pi*phi(j)))...
    .*exp(pi*phi(1:j-1));
%Integrand after the singularity
fU=(ndelta(j+1:end)-ndelta(j))./(exp(pi*phi(j+1:end))-exp(pi*phi(j)))...
    .*exp(pi*phi(j+1:end));
%Compute integrand at singularity
dd=(ndelta(j)-ndelta(j-1))/(phi(j)-phi(j-1))/pi;
%Complete integrand
fI=[fL dd fU];
%Calculate integral
xI=trapezoid(phi,fI);
%Log term from avoiding the singularity
expx=xI+(ndelta(j)/pi)*log(abs((zmax-z(j))/(z(j)-1)));
x(2)=(2*mu/pi)*sqrt(exp(phi(2))-1)*exp(-expx); for j=3:n
    funx=exp(-tau(2:j)).*cos(ndelta(2:j))+x(2);
    x(j)=mu*trapezoid(phi(2:j),funx)+x(2);
    funy=exp(-tau(2:j)).*sin(ndelta(2:j));
    y(j)=mu*trapezoid(phi(2:j),funy);
end
%Plot results
%Plot free surface profile
totaltime=cputime-t1 plot(x,y,ax2,ay2);grid; axis tight;
legend('Numerical Solution', 'Asymptotic Solution'); figure;
%Plot delta
plot(x,ndelta, ax2,adelta2); grid;legend('Numerical Delta',
'Asymptotic Delta'); save newnum1200phi12Fs055.mat x y ndelta phi
Fsp mu; mu Fb=Fsp/mu^1.5

\end{verbatim}

\section{\emph{froude.m}}
\begin{verbatim}
%Equation definition file
function y = froude(guess)
%y is the function to be minimized
%guess is the input vector
%j reprsents equation to use, i represnts variable to perturb
Fsp=0.4; zb=0;
n=length(guess); %Number of equations
mu=guess(1);%mu guess
%Create delta vector
d=[0 guess(2:end)];
%Setting up the zeta and alpha values
phimax=12; phi=linspace(0,phimax,n); z=exp(pi*phi); zmax=z(n); for
j=1:n
    %Calculate the mu value
    if j==1
        nz=500;
        theta=linspace(-pi/2,pi/2,nz);
        zeta=0.5*(sin(theta)+1);
        %Calculate exponential tau
        fun=[];
        for k=1:nz
            if k==nz
                %Avoiding Singularity
                %Improper Integral
                pv=-1*trapezoid(phi(2:n),d(2:n)./...
                    (exp(pi*phi(2:n))-zeta(k)).*exp(pi*phi(2:n)));
                fun(nz)=exp(pv);
            else
                %Improper Integral
                pv=-1*trapezoid(phi,d./(exp(pi*phi)-zeta(k)).*exp(pi*phi));
                fun(k)=exp(pv);
            end
        end
        y(j)=1-(mu/pi)*trapezoid(theta,fun);

    elseif j==n
        %3 point extrapolation
        dend=3*d(n-1)-3*d(n-2)+d(n-3);  %Extrapolated value
        y(j)=dend-d(n);


    else
        %Setting up the y equations
        %First term in equation
        y1=0.5*log(((exp(pi*phi(j))-zb)*(exp(pi*phi(j))-1)...
            /exp(2*pi*phi(j))));

        %Second Term - involes the improper PV integral
        %Integrand before the singularity
        f2L=(d(1:j-1)-d(j))./(exp(pi*phi(1:j-1))-exp(pi*phi(j)))...
            .*exp(pi*phi(1:j-1));
        %Integrand after the singularity
        f2U=(d(j+1:end)-d(j))./(exp(pi*phi(j+1:end))-exp(pi*phi(j)))...
            .*exp(pi*phi(j+1:end));
        %Compute integrand at singularity
        dd=(d(j)-d(j-1))/(phi(j)-phi(j-1))/pi;

        %Complete integrand
        f2I=[f2L dd f2U];
        %Calculate integral
        y2I=trapezoid(phi(1:end),f2I(1:end));
        %Log term from avoiding the singularity
        y2=y2I+(d(j)/pi)*log(abs((zmax-z(j))/(z(j)-1)));

        %Third term - involves integral
        const=-3*mu^3/Fsp^2;
        y3=(-1/3)*log(abs(const*trapezoid(phi(1:j),sin(d(1:j)))));
        y(j)=y1+y2+y3;
    end
end
\end{verbatim}

\section{\emph{trapezoid.m}}
\begin{verbatim}
%Trapezoid Scheme
function z = trapezoid(x,y) nx=length(x); ny=length(y); if nx==ny
    if nx==1
        z=0;
    else
        z = diff(x).* (y(1:ny-1) + y(2:ny))/2; %Area for each trapezoid
        z=sum(z);   %Sum trapezoids
    end
else
    'Vector lengths do not match'
end
\end{verbatim}

\bibliographystyle{unsrt}



\begin{thebibliography}{10}

\bibitem{H1995}
G.C. Hocking.
\newblock Supercritical withdrawal from a two-layer fluid through a line
  source.
\newblock {\em Journal of Fluid Mechanics}, 297:37--47, March 1995.

\bibitem{HF1997}
G.C. Hocking and L.K. Forbes.
\newblock Withdrawal from a two-later invisid fluid in a duct.
\newblock {\em Journal of Fluid Mechanics}, 361:275--296, December 1997.

\bibitem{T1991}
E.O. Tuck.
\newblock Ship-hydrodynamics free-surface problems without waves.
\newblock {\em Journal of Ship Research}, 35(4):277--287, 1991.

\bibitem{V1997}
J.-M. Vanden-Broeck.
\newblock Numerical calculations of the free-surface flow under a sluice gate.
\newblock {\em Journal of Fluid Mechanics}, (330):339--347, 1997.

\bibitem{STOKER}
J.J. Stoker.
\newblock {\em Water Waves: The Mathematical Theory with Applications}.
\newblock Interscience Publishers Inc., 1st edition, 1957.

\bibitem{DEBNATH}
L.~Debnath.
\newblock {\em Nonlinear Water Waves}.
\newblock Academic Press, 1st edition edition, 1994.

\bibitem{HF1992}
G.C. Hocking and L.K. Forbes.
\newblock Subcritical free-surface flow caused by a line source in a fluid of
  finite depth.
\newblock {\em Journal of Engineering Mathematics}, (26):455--466, 1992.

\bibitem{MV1991}
H.~Mekias and J.-M.~Vanden Broeck.
\newblock Subcritical flow with a stagnation point due to a soruce beneath a
  free surface.
\newblock {\em Phys. Fluids A}, (11):2652--2658, 1991.

\bibitem{V1996}
J.-M. Vanden-Broeck.
\newblock Waves generated by a source below a free surface in water of finite
  depth.
\newblock {\em Journal of Engineering Mathematics}, (30):603--609, 1996.

\bibitem{HF2000}
G.C. Hocking and L.K. Forbes.
\newblock Withdrawal from a fluid of finite depth through a line sink,
  including surface-tension effects.
\newblock {\em Journal of Engineering Mathematics}, (38):91--100, 2000.

\bibitem{Thomson}
L.M.~Milne Thomson.
\newblock {\em Theoretical Hydrodynamics}.
\newblock MacMillan, 5th edition, 1974.

\bibitem{B1999}
J.P. Boyd.
\newblock The devil's invention: asymptotic, superasymptotic and
  hyperasymptotic series.
\newblock {\em Acta Applicandae Mathematicae: An International Survey Journal
  on Applying Mathematics and Mathematical Applications}, 56:1--98, March 1999.

\bibitem{BATCHELOR}
G.K. Batchelor.
\newblock {\em An Introduction to Fluid Dynamics}.
\newblock Cambridge University Press, 1st edition, 1967.

\bibitem{BOAS}
M.~L. Boas.
\newblock {\em Mathematical Methods in the Physical Sciences}.
\newblock Wiley, 2nd edition, 1983.

\bibitem{RAHMAN}
M.~Rahaman.
\newblock {\em Water Waves: Relating Modern Theory to Advanced Engineering
  Practice}.
\newblock Oxford Science Publications, 1st edition, 1995.

\bibitem{Arfken}
G.B. Arfken and H.~J. Weber.
\newblock {\em Mathematical Methods for Physicists}.
\newblock Academic Press, 4th edition, 1995.

\bibitem{WL}
J.V. Wehausen and E.V. Laitone.
\newblock Surface waves.
\newblock In {\em Handbuch der Physik}.

\bibitem{NUMR}
W.~T.~Vetterling W.H.~Press, S.A.~Teukolsky and B.~P. Flannery.
\newblock {\em Numerical Recipes in C}.
\newblock Cambridge Press, 2nd edition, 1997.

\bibitem{HV1997}
G.C. Hocking and J.-M. Vanden-Broeck.
\newblock Withdrawal of a fluid of finite depth through a line sink with a cusp
  in the free surface.
\newblock {\em Computers and Fluids}, 27:797--806, December.

\bibitem{MATLAB}
M.A.~Branch T.~Coleman and A.~Grace.
\newblock {\em Optimization Toolbox User's Guide}.
\newblock The Maths Works Inc., third edition edition, 1999.

\bibitem{MORE}
J.~J. Mor{\'e}.
\newblock {\em Numerical Analysis}.
\newblock Springer Verlag, 1st edition, 1977.

\bibitem{abram}
M.~Abramowitz and I.~Stegun.
\newblock {\em Handbook of Mathematical Functions}.
\newblock Dover, 1970.

\bibitem{CV2002}
S.J. Chapman and J.-M. Vanden-Broeck.
\newblock Exponential asymptotics and capillary waves.
\newblock {\em SIAM Journal of Applied Mathematics}, 62:1872--1898, 2002.

\bibitem{M1988}
M.A.D. Madurasinghe.
\newblock Splashless ship bows with stagnant attachment.
\newblock {\em Journal of Ship Research}, 32:194--202, 1988.

\bibitem{FT1995}
D.E. Farrow and E.O. Tuck.
\newblock Further studies of stern wakemaking.
\newblock {\em Journal of the Australian Mathematical Society, Series B},
  36:424--437, 1995.

\bibitem{MF1999}
S.W. McCue and L.K. Forbes.
\newblock Bow and stern flows with constant vorticity.
\newblock {\em Journal of Fluid Mechanics}, 399:277--300, 1999.

\bibitem{VT1977}
J.-M. Vanden-Broeck and E.O. Tuck.
\newblock Computation of near-bow or stern flows, using series expansion in the
  froude number.
\newblock {\em Proceedings of the 2nd International Conference of Numerical
  Ship Hydrodynamics, Berkeley}, pages 371--381, 1977.

\bibitem{FS1968}
D.D. Fangmeier and T.S. Strelkoff.
\newblock Solution for gravity flow under a sluice gate.
\newblock {\em Journal of the Engineering Mechanics Division ASCE},
  94:153--176, 1968.

\end{thebibliography}

\end{document}